\documentclass[twocolumn]{aa}

\usepackage{graphicx}
\usepackage{txfonts}
\usepackage[dvipsnames]{xcolor}
\usepackage{soul}
\usepackage{hyperref}
\usepackage[capitalise]{cleveref}
\usepackage{siunitx}
\usepackage{gensymb}
\usepackage[normalem]{ulem}

\hypersetup{
    colorlinks,
    linkcolor={red!50!black},
    citecolor={blue!50!black},
    urlcolor={blue!80!black}
}

\DeclareSIUnit \year {yr}
\DeclareSIUnit \parsec {pc}
\DeclareSIUnit \eV {eV}
\DeclareSIUnit \Msun {M_\odot}

\begin{document}

   \title{The Sagittarius stellar stream embedded in a fermionic dark matter halo}

   \author{Santiago Collazo\inst{1,2}\fnmsep\thanks{
         \email{scollazo@fcaglp.unlp.edu.ar}}
          \and
          Martín F. Mestre\inst{1,2}
          \and
          Carlos R. Argüelles\inst{1,3}}

   \institute{Instituto de Astrof{\'i}sica de La Plata (CONICET-UNLP), Paseo del Bosque S/N, La Plata (1900), Buenos Aires, Argentina
        \and
            Facultad de Ciencias Astron{\'o}micas y Geof{\'i}sicas de La Plata (UNLP), Paseo del Bosque S/N, La Plata (1900), Buenos Aires, Argentina
        \and
            ICRANet, Piazza della Repubblica 10, 65122 Pescara, Italy}

   \date{}

\abstract{
    Stellar streams are essential tracers of the gravitational potential of the Milky Way, with key implications to the problem of dark matter model distributions, either within or beyond phenomenological $\Lambda$CDM halos.
}
{
    For the first time in the literature, a dark matter (DM) halo model based on first physical principles such as (quantum) statistical mechanics and thermodynamics is used to try to reproduce the $6$D observations of the Sagittarius (Sgr) stream. Thus, we aim to extract quantitative and qualitative conclusions on how well our assumptions stand with respect to the observations. We model both DM haloes, the one of the Sgr dwarf and the one of its host, with a spherical self-gravitating system of neutral fermions which accounts for the effects of particles escape and fermion degeneracy (due to the Pauli exclusion principle), the latter causing a high-density core at the center of the halo. Full baryonic components for each galaxy are also considered.
}
{
   We use a spray algorithm with $\sim10^5$ particles to generate the Sgr tidal debris, which evolves in the combined gravitational potential of the host-progenitor system, to then make a direct comparison with the full phase-space data of the stream. We repeat this kind of simulations for different parameter setups of the fermionic model including the particle mass, with special attention to test different DM halo morphologies allowed by the physics, including polytropic density tails as well as power-law-like trends.
}
{
   We find that across the different families of fermionic halo models, they can only reproduce the trailing arm of the Sgr stream. Within the observationally allowed span of enclosed masses where the stream moves, neither the power-law-like nor the polytropic behavior of the fermionic halo models can answer for the observed trend of the leading tail. A conclusion which is shared by former analysis using other types of spherically symmetric haloes. Thus, we conclude that further model sophistications, such as abandoning spherical symmetry and including the Large Magellanic Cloud perturber, are needed for a proper modelisation of the overall Milky Way potential within this kind of first-principle halo models.
}
{
}

   \titlerunning{}

   \maketitle

\section{Introduction}

Stellar streams are important tracers of the gravitational potential of the Galaxy, extending from inner to outermost radial extents, thus allowing to infer the mass distribution of the host \citep[see e.g.][for the current state of tidal debris in the Milky Way]{Li2019, Mateu2023, Ibata2024, Bonaca2024}.

The information content on the gravitational potential of the Milky Way (MW) provided by cold stellar streams, such as the iconic GD-1~\citep{Grillmair_2006},  was theoretically studied by~\cite{2018ApJ...867..101B}. Among a series of interesting results, they found that in simple analytic models of the MW, those streams in eccentric orbits are better for constraining the halo shape. Generally, for each single stream they find degeneracies between the properties of the dark matter halo and the baryonic potentials. They show that simultaneous fit of multiple streams will constrain all parameters at the percent level using basis function expansions to give the gravitational potential more freedom.

Within the vast spectra of tidal stream properties, there exists in the MW examples of long stellar streams like the Orphan-Chenab and the Sagittarius streams. Their currently measured angle subtended in the sky are  $210^\circ$ \citep{Koposov2019} and more than $360^\circ$ \citep{Majewski2003, Ibata2020}, respectively.  The former stream has a width of $\sim 2^\circ$, being rather thin in spite of having an unknown massive progenitor \citep{Hawkins2023, Belokurov2007}. In this work we study the latter stream, which was tidally built from the Sgr dwarf progenitor. It is the most extended in length and width from all the MW streams observed up to date, and it is possible to get crucial information about the dark matter halo potential and morphology \citep{Ibata2001, Helmi2004, Law2005, Johnston2005, Law2009, Law2010, Deg2013, Vasiliev2021}.

The first observations of the Sgr stream progenitor was performed by \cite{Ibata1994}. After that, a big effort was made to determine the remnants structures arising from this progenitor \citep[e.g.,][]{Totten1998, Mateo1998, Majewski1999, Dinescu2002,  Newberg2002, Newberg2003, Newberg2007, Yanny2009, Koposov2012, Koposov2013, Slater2013, Huxor2015, Luque2017, Hernitschek2017, Antoja2020, Ramos2020, Ibata2020}. The first global spatial characterization of the stellar stream stripped from Sagittarius was obtained by \cite{Majewski2003} with the Two Micron All Sky Survey (2MASS) using M giant stars data. In such observation, it was concluded that the tidal debris left by this satellite develops two tails: the leading one spreading along the northern Galactic hemisphere, and the trailing one lying in the southern part of it. Afterwards, \cite{Majewski_2004AJ} performed radial velocity measurements, enabling subsequent dynamical modeling of the stream~\citep{Law2005}.

The structure and evolution of the Sgr dwarf stream, together with the overall gravitational potential of the host Galaxy, are still open, coupled problems because there is not yet a single full model able to fit the astrometric, photometric, and spectrometric observations performed up to date. Successful enough examples in that direction were done by~\cite{Law2010} through $N-$body simulations, modeling the tidal disruption suffered by the Sgr dwarf on its way around the Milky Way. For that purpose, they fitted radial velocity data of a sample of 2MASS M-giants in combination with SDSS photometry. Similar results were obtained in \cite{Deg2013}, where there was performed a Bayesian analysis using single-particle orbits to model the Milky Way potential. In \cite{Gibbons2014}, the predicted Sgr tidal stream was used to measure the cumulative mass of the Galaxy based on the values of the apocentric distances of the arms and the precession angle of the stream \citep{Belokurov2014}. In \cite{Fardal2019}, they tested their kinematics and spatial predictions of Sgr tidal debris with RR Lyrae observations from Pan-STARRS1, paying special attention to the apocenter distances of the stream. The work of \cite{Dierickx2017} has proved the main features of the Sgr stream using mock tidal debris computed with GADGET \citep{Springel2005}. Further improvements in the stream N-body model have also been made by~\cite{2022ApJ...940L...3W} with a spherical DM halo and by~\cite{Vasiliev2021} through the inclusion of the Large Magellanic Cloud (LMC). The latter model constitutes the paradigm that agrees best with current observations of the Sgr stream.

Despite the reliability of the numerical simulations, it is necessary to improve the large computation times and to try semi-analytical methods in order to be more flexible to study different DM halo models and their associated free parameter space. This can be achieved by the use of ejection algorithms (sometimes called spray methods), like the ones implemented in \cite{Varghese2011}, \cite{Kupper2012}, \cite{Gibbons2014}, \cite{Bowden2015}, and \cite{Fardal2015}. The methodology of these algorithms consists of the ejection of stars with adequate initial conditions at the two Lagrange points around the progenitor, to then integrate them over a large enough time-span and get the orbit of all the stars. As demonstrated in \cite{Gibbons2014}, in order for this semi-analytical method to be almost indistinguishable from the simulations, it is needed to include the action of the gravity of the progenitor in addition to that of the host.

Most studies aimed to reproduce the Sgr stream were developed under the assumption of phenomenological DM halo profiles. For instance, by using different axisymmetric halo potentials such as prolate or oblate logarithmic functions \citep{Johnston2005, Law2005}, by applying axisymmetric variations of traditional $\Lambda$CDM halos \citep{Law2005}, or by employing very specific realizations of triaxial halos \citep{Law2010}\footnote{In \cite{Debattista2013} it has been proved the difficulty of reconciling the \cite{Law2010} model (where the minor axis of the baryonic disk is aligned with the intermediate axis of the DM halo) with CDM halo predictions due to an instability issue.}. Maybe the most notorious example is the DM halo model suggested by \citep{Vasiliev2021}, where they considered a complex halo whose axis ratios and orientation of its principal axes vary with radius. Their best-fit model has two effective components, one rotated with respect to the other, in addition to the perturbative effects of the LMC. However, the use of such purely phenomenological DM profiles is impractical for obtaining information about the underlying DM nature, particle mass, and cosmological framework. An interesting, though poorly investigated alternative is to study stellar streams via numerical methods which incorporate DM halo potentials obtained from first-principle physics. Such models are based on statistical mechanics and thermodynamics, in which the quantum nature of the DM particles is included. This has been studied in \cite{Robles2015} for bosonic DM particles, where the DM halo is modeled via a self-gravitating system of self-interacting scalar-field dark matter (SFDM), which forms a Bose-Einstein condensate. However, that work was only applied to hypothetical (or generic) dwarf satellites hosted by a Milky Way-like galaxy, and mainly focused in comparing the tidal effects with respect to CDM halos.

Recently, \cite{Mestre2024} have modeled the {GD-1} 5D track with a MW that includes a fermionic DM halo model built from first principles, such as the Ruffini-Arg\"uelles-Rueda (RAR) model (see \citealt{Ruffini2015,Arguelles2018} for the original works, and \citealt{2023Univ....9..197A} for a review). The RAR model is built in terms of a self-gravitating system of massive fermions at finite temperature, whose more general solutions can include two different regimes in the same system. One to be highly degenerate at the center forming a dense fermion core (thanks to the Pauli exclusion principle) and one diluted in the halo region (resembling the King profile). The successful phenomenological analysis to this model has been recently shown for different galaxy types and scaling relations \citep{Arguelles2019,2023ApJ...945....1K}, including our own Galaxy \citep{Arguelles2018,BecerraVergara2020,2021MNRAS.505L..64B,2022MNRAS.511L..35A,2023Univ....9..372A}. In the latter case, we showed that the central DM core can work as an alternative to the massive BH in SgrA*, while the outer halo explains the Galaxy rotation curve.

Important aspects of this kind of first-principle physics models (for bosonic or fermionic DM) with respect to the ones arising from classical N-body simulations, is that the resulting density profiles are of cored\footnote{By the term ``cored'' we refer to the fact that the profiles present a ``plateau'' at inner halo scales. Not to be confused with the degenerate and compact core at milliparsec scales of the core--halo distribution, like the model here studied. Our model has both a central compact core and a plateau at the halo.} nature at inner halo scales (i.e. not cuspy-like as in CDM-only cosmologies), and that do depend on the particle mass (see e.g. \cite{2021MNRAS.502.4227A} for the fermionic case). Other relevant advantages of this kind of astrophysical core--halo fermionic DM profiles, is that one can apply thermodynamical tools and demonstrate that they remain stable during cosmological timescales once formed \citep{2021MNRAS.502.4227A}. Also, one can show that it exists a critical fermion core mass above which it collapses towards an SMBH with important applications to the high redshift Universe \citep{2023MNRAS.523.2209A,2024ApJ...961L..10A}.

Thus, in this work we aim to reproduce, for the first time in the literature, the main features of the Sgr stellar stream observations, assuming that both the DM halo of our Galaxy and the one of the Sgr dwarf are composed of fermionic dark matter halos according to the RAR model \cite{Arguelles2018}.  For such an endeavor, we will apply the ejection method as proposed in \cite{Gibbons2014}, and thus we will consider the self-gravity caused by the satellite. This will allow us to explore a considerable range of free parameters of the DM models in manageable computing times, paying special attention to test different DM halo tails allowed by the physics of the fermionic model.

The flow of the work is as follows. In Sec. \ref{sec:observations} we detail the data of the Sgr dwarf stream used to compare with the theoretical predictions. In Sec. \ref{sec:potential}, we explain the different mass distributions assumed to model the gravitational potentials of both the Galaxy and the satellite. In Sec. \ref{sec:stream_generation}, it will be introduced the spray method, together with the specific tidal radius approximation which are going to depict the stellar stream. In Sec. \ref{sec:comparison} we show the comparison between the tidal debris predicted by our procedure with the observed data (including proper motions, line of sight velocity, distance, and projected spatial coordinates). Finally, in Sec. \ref{sec:final} we draw our conclusions.

\section{Stream Observables}\label{sec:observations}

The observed properties of the Sgr stream are taken from \cite{Vasiliev2021} and \citet{Ibata2020} (hereafter I$+$2020); both datasets based on Gaia DR2 and the former also in the 2MASS catalog. In I$+$2020, the STREAMFINDER algorithm \citep{STREAMFINDER2018} was there used to search for candidate stars tidally stripped from the progenitor. Their analysis gave a map of stars with a detection significance $>15\sigma$ (note that this statistic implies that each star has position and velocity compatible with stream-like behavior but does not certainly imply stream membership). Those selections of stars were represented in the heliocentric Sgr coordinates system $(\Lambda_{\odot}, B_{\odot})$ developed by \cite{Majewski2003}, although it was chosen the approach of \cite{Koposov2012} of inverting the latitude axis, $B_{\odot}$. The equatorial plane of this system is defined as the one whose north pole has Galactocentric coordinates $(l,b)=(93.8\degree, 13.5\degree )$. On the other hand, the longitudes are defined to increase in the direction of the trailing debris, with $\Lambda_{\odot} = 0\degree$ chosen to be the longitude of the center of the King profile associated with the main body of Sgr as done by \cite{Majewski2003}. This is the reference system we are using in this study to contrast predictions with observations.

The work of \cite{Vasiliev2021} is based mainly on red giant branch (RGB) stars to get proper motions, line of sight velocities, and spherical coordinates of stream candidates. To complete the set of observations and make it a full 6D dataset, they used RR Lyrae stars to extract estimations of the distance trend along the tidal arms and the remnant. The contaminants were filtered out through a series of cuts in proper motions, stream width, color, and magnitude, yielding $\sim 55000$ stars as a final sample.

A Galactocentric solar distance of $R_{\odot} = 8.122 \pm 0.031$ kpc \citep{GravityCollaboration2018} was chosen. Based on \cite{Reid2014} we assume $v_{\mathrm{c}}(R_{\odot}) + V_{\mathrm{LSR,pec}} + V_{\odot} = 255.2 \pm 5.1$ km/s, where $V_{\mathrm{LSR,pec}}$ is the $V$ component of the peculiar velocity of the LSR with respect to the Galactic center, $V_{\odot}$ is the peculiar velocity of the Sun with respect to the LSR and $v_{\mathrm{c}}(R_{\odot})$ is the circular velocity at the location of the Sun, which is model dependent. The other two components for the peculiar velocity of the Sun, $U_{\odot}$ and $W_{\odot}$, were taken from \cite{Schonrich2010}, and respectively read $U_{\odot} = 11.1$ km/s and $W_{\odot} = 7.25$ km/s. We assume that the LSR has zero $U$ and $W$ velocity components.

\begin{figure}
    \centering
    \begin{tabular}{@{}c@{}}
        \includegraphics[width=\columnwidth]{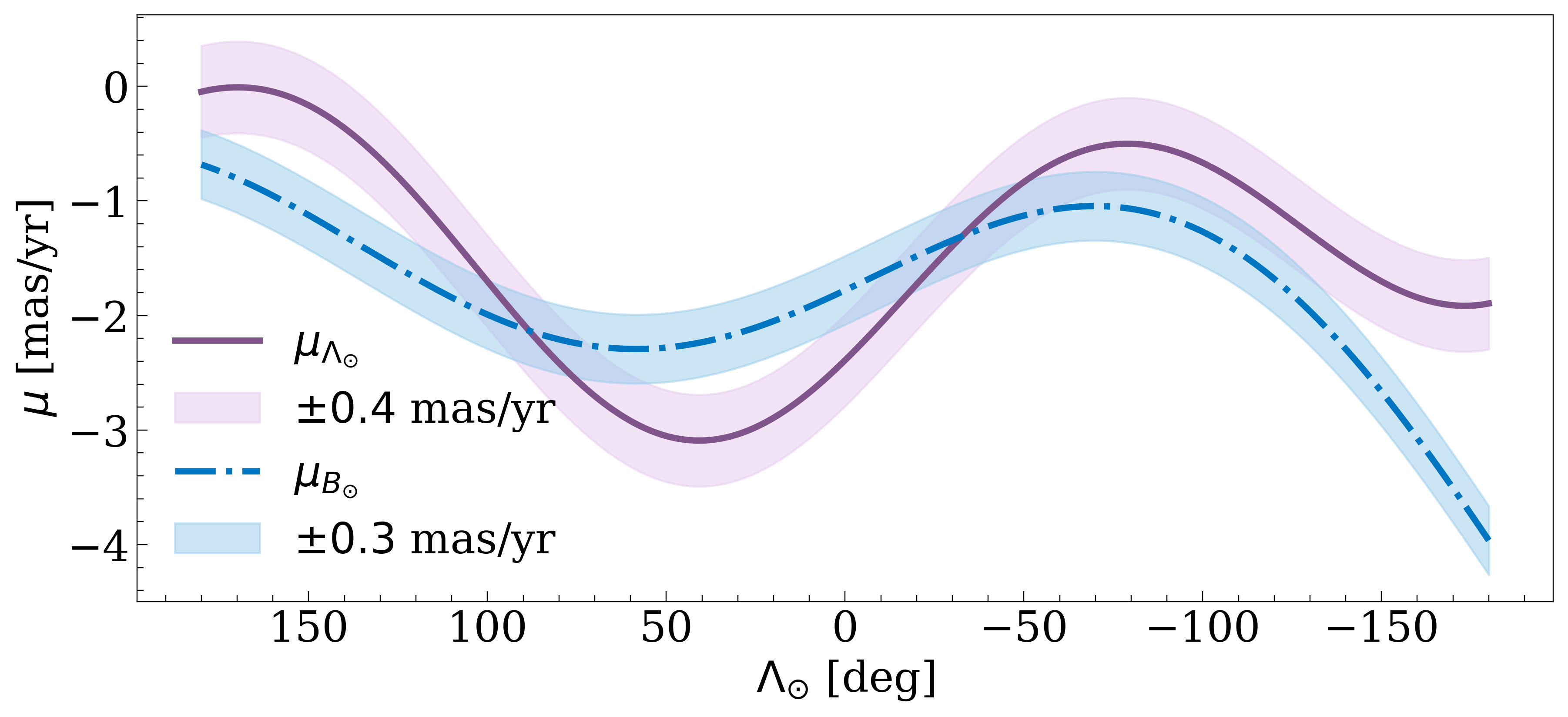}
    \end{tabular}
    \begin{tabular}{@{}c@{}}
        \includegraphics[width=\columnwidth]{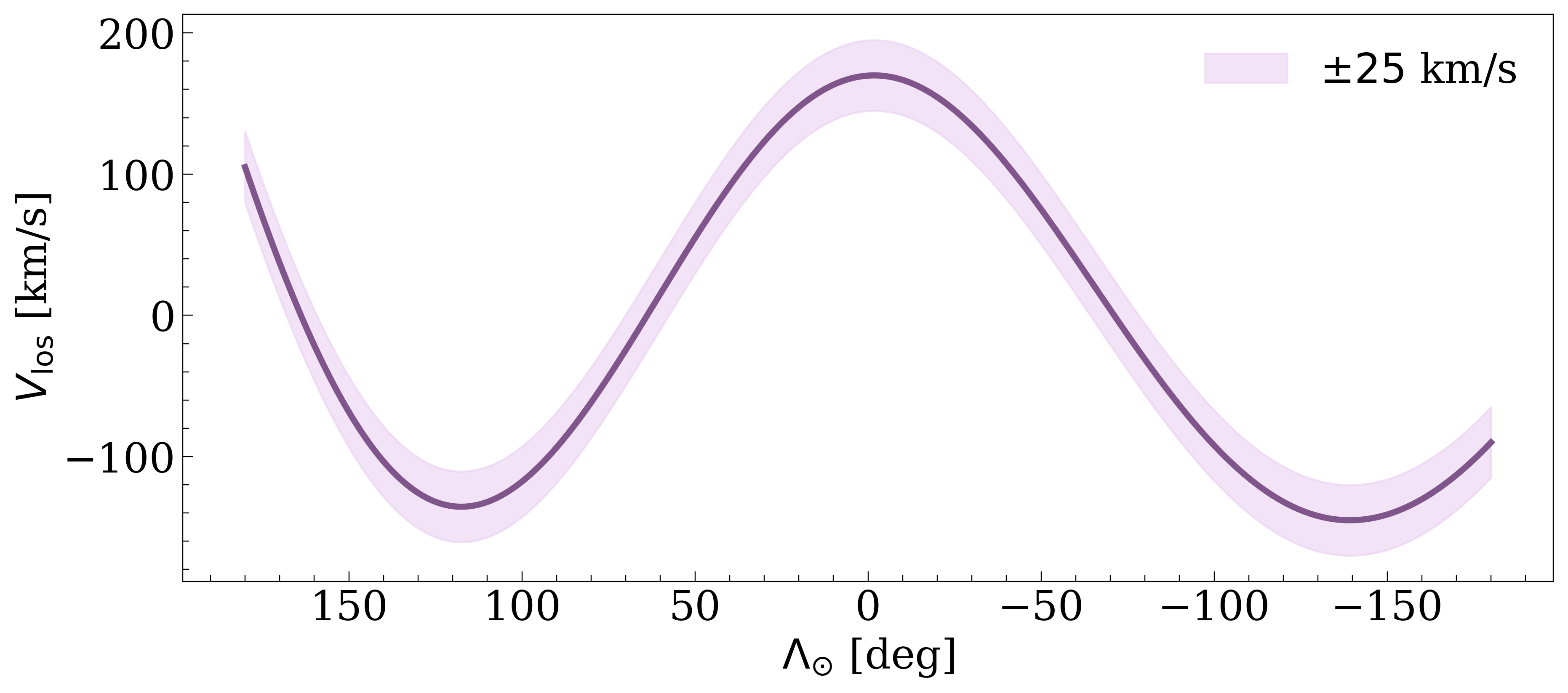}
    \end{tabular}
    \begin{tabular}{@{}c@{}}
        \includegraphics[width=\columnwidth]{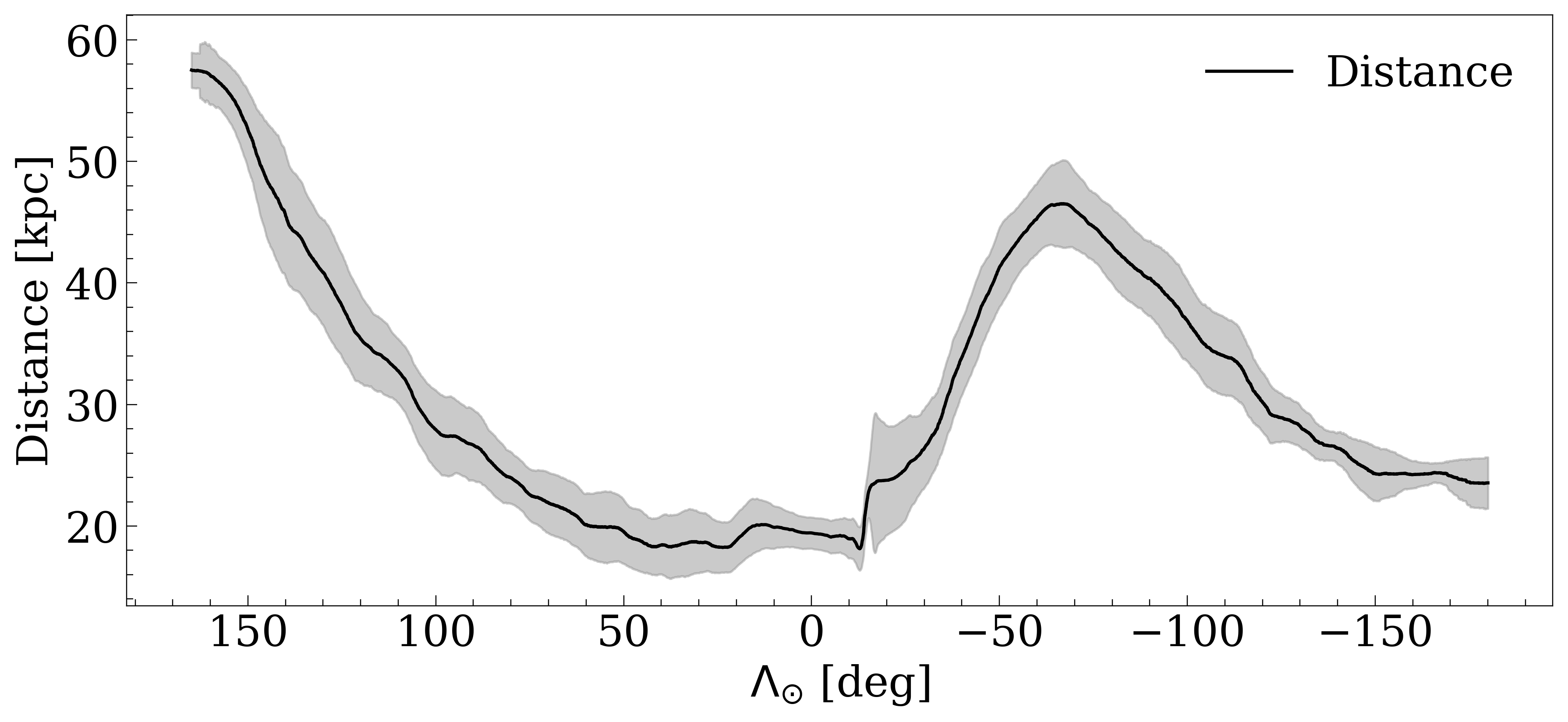}
    \end{tabular}
    \caption{Best-fit proper motions (upper panel) and line of sight velocity (middle panel) functions described by the expressions given in Eqs. (\ref{eq:fitted_function}) and (\ref{eq:los_vel}), respectively. Their coefficients are shown in Table \ref{tab:coefficients_fit}. The shaded regions in the proper motion plot show the confidence intervals (see Figs. 2b and 2c of \cite{Ibata2020} to contrast the polynomials with the stars distribution) such that the stream stars selected by I$+$2020 have a contamination of $11\%$. The shaded region in the radial velocity plot corresponds to half the interval that selects stars with a global contamination of $18\%$ according to I$+$2020. Bottom panel: moving average Galactocentric distance as a function of the longitude $\Lambda_{\odot}$ for the set of stars given in \cite{Vasiliev2021}, with the grey shaded region representing the corresponding moving average error along the stream.}
    \label{fig:mu_vs_lambda}
\end{figure}

Thus, the complete set of observations we are using is a full 6D map of stars composed by the ones processed in I$+$2020 for the momentum space and in \cite{Vasiliev2021} for the configuration space. To characterize the latter, we use the cartesian coordinates of the stars whose distance trend was estimated using the subset of RR Lyrae stars as specified in \cite{Vasiliev2021}. They allow to display the trend of the spatial distribution of the stars in the leading and trailing arms. We will use the projection of the positions in the $x-z$ plane and the distance of those stars as a function of $\Lambda_{\odot}$. The debris distance is computed as a moving mean as a function of $\Lambda_{\odot}$, and is displayed on the bottom panel of Fig.~\ref{fig:mu_vs_lambda}.

Regarding the velocity space, the observations are the two proper motions of the stars ($\mu_{\Lambda_{\odot}}$, $\mu_{B_{\odot}}$)\footnote{$\mu_{\Lambda_{\odot}}$ includes the $\cos(B_{\odot})$ factor throughout this work.} and the line of sight velocity. The number of stars filtered by I$+$2020 is about $\sim2\times 10^5$, and they fitted analytical functions to the corresponding distributions of stars in the three planes, $\Lambda_{\odot}-\mu_{\Lambda_{\odot}}$, $\Lambda_{\odot}-\mu_{B_{\odot}}$, and $\Lambda_{\odot}-V_{\mathrm{los}}$. Next, we show the generic formula used for both proper motions (see Table \ref{tab:coefficients_fit} for the corresponding coefficients in each case):

\begin{equation}
    \mu(\Lambda_{\odot}) = a_{1}\mathrm{sin}(a_{2}\Lambda_{\odot} + a_{3}) + a_{4} + a_{5}\Lambda_{\odot} + a_{6}\Lambda_{\odot}^{2},
    \label{eq:fitted_function}
\end{equation}
where $\Lambda_{\odot}$ is in degrees. The function fitted to the distribution of stars in the $\Lambda_{\odot}-V_{\mathrm{los}}$ plane is (see Table \ref{tab:coefficients_fit}):

\begin{equation}
    V_{\mathrm{los}}(\Lambda_{\odot}) = a_{1}\mathrm{cos}\left(a_{2} + a_{3}t + a_{4}t^{2} + a_{5}t^{3}\right) + a_{6} + a_{7}\Lambda_{\odot} + a_{8}\Lambda_{\odot}^{2},
    \label{eq:los_vel}
\end{equation}
where $t = \Lambda_{\odot}\pi/180\degree$.

\begin{table}
    \caption{Fitted coefficients.}
    \centering
    \begin{tabular}{cccc}
        \hline
        Coefficient & $\mu_{\Lambda_{\odot}}$ & $\mu_{B_{\odot}}$ & $V_{\mathrm{los}}$ \\
        \hline
        $a_{1}$ & $1.1842$ & $-1.2360$ & $277.4481$ \\
        $a_{2}$ & $-1.5639\pi /180^{\circ}$ & $1.0910\pi /180^{\circ}$ & $0.1088$ \\
        $a_{3}$ & $-0.39917$ & $0.3633$ & $1.1807$ \\
        $a_{4}$ & $-1.9307$ & $-1.3412$ & $0.0400$ \\
        $a_{5}$ & $-8.0606\cdot 10^{-4}$ & $7.3022\cdot 10^{-3}$ & $-0.0145$ \\
        $a_{6}$ & $3.2441\cdot 10^{-5}$ & $-4.3315\cdot 10^{-5}$ & $-106.0409$ \\
        $a_{7}$ & $-$ & $-$ & $0.4529$ \\
        $a_{8}$ & $-$ & $-$ & $0.0110$ \\
        \hline
    \end{tabular}
    \tablefoot{Coefficients of the different functions given in Eqs. (\ref{eq:fitted_function}) and (\ref{eq:los_vel}) used to fit the distributions of proper motion and line of sight velocities of stars as a function of the longitude $\Lambda_{\odot}$.}
    \label{tab:coefficients_fit}
\end{table}

The upper panel of Fig.~\ref{fig:mu_vs_lambda} shows both proper motions described by the formulas above. It is seen that the proper motion has a sinusoidal component on the interval $-180^{\circ} \leq \Lambda_{\odot} \leq 180^{\circ}$, which corresponds to a whole wrap around the Milky Way. The same behavior can be seen in the middle panel of Fig.~\ref{fig:mu_vs_lambda}, which displays the line of sight velocity trend of the stream stars. Additionally, the galactocentric distance of the stream stars as a function of the longitude $\Lambda_{\odot}$ (bottom panel of Fig. \ref{fig:mu_vs_lambda}) helps to represent the whole wrap of the stream around the Galaxy. This will be shown explicitly in the comparison between the predictions and the observations in the galactocentric $x-z$ plane in Sec. \ref{sec:comparison}.

It is well known that the debris of the progenitor goes all the way around the Galaxy, as shown e.g. in \cite{Vasiliev2021} through their predictions and compared to data corresponding to the second wrap of the trailing arm or in \cite{Hernitschek2017} through RR Lyrae stars distribution, depicted in their Fig.~4. Following \cite{Ibata2020} and \cite{Vasiliev2021}, we will consider only stars within $180\degree$ at each side of the progenitor to compare with the observations on the phase space, thus neglecting older wraps.

\section{Gravitational potentials and progenitor orbit}\label{sec:potential}

\subsection{Milky Way potential}
\label{sec:MWPot}

The strategy to fix the Galactic potential on which the spray simulations will be performed is inspired by the work developed in \cite{Gibbons2014}. From that work, it is possible to obtain the accumulated (total) mass of the host at different radii, such that it maximizes the likelihood of having the required (i.e observed) apocentric Galactocentric distances of the Sgr stream's arms and their angular difference. They used measurements from \cite{Belokurov2014}, which are approximately $a_{L}=47.8\pm 0.5$~kpc for the leading arm apocenter, $a_{T}=102.5\pm 2.5$ kpc for the trailing one, and $\psi = 99.3\degree \pm 3.5\degree$ for the (heliocentric) opening angle between both given apocenters. These values are in agreement with those given by \cite{Hernitschek2017}, who measures $a_{L} = 47.8\pm0.5$~kpc, $a_{T} = 98.95\pm1.3$~kpc and $\psi = 104.4\degree \pm 1.3\degree$. The corresponding accumulated Galaxy masses are reported in Fig.~13 and Table 3 of \cite{Gibbons2014} from which we will choose a 3-tuple of DM-mass values (i.e. once extracted the baryonic component, see below) for each DM potential. Each tuple is conformed of the enclosed DM mass at three different characteristic scales where the stream moves: inner ($r=12$ kpc), mid ($r=40$ kpc), outer ($r=80$ kpc), shown in~\cref{tab:accumulated_mass}. These selected masses to be fulfilled by the DM models belong to $2 \sigma$ mass dispersion interval reported in Fig.~13 of \cite{Gibbons2014}.

Given the large amount of ejected stars ($\sim 10^5$), it takes about $40$ hours of CPU time within our Python code to predict one Sgr stream, precluding us from running a Monte Carlo algorithm on the model parameters. This is why we will present only four different simulations, with a conveniently fixed DM model each: three different RAR models with different degree of accumulated mass from inner to outer halo Galaxy scales, and a Burkert model for comparison (see below for details). On the other hand, the baryonic component of the Galaxy will be modeled as the combination of three main components: a bulge and two disks (thick and thin), based on \cite{Pouliasis2017}. The bulge is described through a Plummer sphere while the thin and thick disks are modeled by the Miyamoto-Nagai formula \citep{Miyamoto1975}. Such baryonic models have gravitational potentials of the form:

\begin{equation}
    \Phi_{\mathrm{MN}}(R,z) = - \frac{GM}{\sqrt{R^{2} + \left(a + \sqrt{z^{2} + b^{2}}\right)^{2}}},\label{eq:MN}
\end{equation}

\begin{equation}
    \Phi_{\mathrm{P}}(r) = - \frac{GM}{\sqrt{r^{2} + b^{2}}},\label{eq:P}
\end{equation}
where MN stands for Miyamoto-Nagai and P for Plummer. The values of the free parameter for each baryonic component are given in Table~\ref{tab:coefficients_baryons}.

\begin{table}
    \caption{Coefficients of the Galactic baryonic components.}
    \centering
    \begin{tabular}{cccc}
        \hline
        Parameter & Bulge & Thin disk & Thick disk \\
        \hline
        $M$ [$M_{\odot}$]& $1.07\times10^{10}$ & $3.94\times10^{10}$ & $3.94\times10^{10}$ \\
        $a$ [kpc] & $-$ & $5.3$ & $2.6$ \\
        $b$ [kpc] & $0.3$ & $0.25$ & $0.8$ \\
        \hline
    \end{tabular}
    \tablefoot{Table adapted from Table 1, column corresponding to Model I, of \cite{Pouliasis2017}.}
    \label{tab:coefficients_baryons}
\end{table}

\begin{table*}
    \caption{Mass values of the host.}
    \centering
    \begin{tabular}{ccccccc}
        \hline
        M(r) [$M_\odot$] & RAR $1$ ($m=56$ keV) & RAR $2$ ($m=56$ keV) & RAR $3$ ($m=20$ keV) & Burkert  \\
        \hline
        $M(r_c$) & $3.5\times10^6$ & $3.5\times10^6$ & $3.5\times10^6$ & $-$ \\
        $M(r=12~\mathrm{kpc}$) & $3\times10^{10}$ & $7\times10^{10}$ & $5\times10^{10}$ & $5\times10^{10}$ \\
        $M(r=40~\mathrm{kpc}$) & $2.2\times10^{11}$ & $2.5\times10^{11}$ & $2.2\times10^{11}$ & $2.2\times10^{11}$ \\
        $M(r=80~\mathrm{kpc}$) & $2.4\times10^{11}$ & $2.7\times10^{11}$ & $4\times10^{11}$ & $4\times10^{11}$ \\
        \hline
    \end{tabular}
    \tablefoot{Accumulated DM mass values used to fit the RAR 2, RAR 3, and Burkert halos of the hosts involved in the spray algorithm. The values are taken from \cite{Gibbons2014}, as detailed in Sec. \ref{sec:potential}. $r_{c}$ is the core radius of the fermionic models. While both the models RAR 1 and RAR 2 develop a polytropic halo tail with a dense DM-core (on miliparsec scales) alternative to the central BH, the model RAR 3 develops a power-law-like halo trend (similar to Burkert).}
    \label{tab:accumulated_mass}
\end{table*}

Regarding the DM component, we will follow the RAR model~\citep{Arguelles2018}, sometimes referred to in the literature as the relativistic fermionic King model. This is a semi-analytical model based on a self-gravitating system of neutral fermions, derived from first physical principles. It is based on the knowledge of a most likely coarse-grained phase-space distribution function (DF) of the fermions at (violent) relaxation. It is obtained by applying a Maximum Entropy Production Principle on an adequate kinetic theory coupled with gravity \citep{Chavanis2}. Such a DF is of Fermi-Dirac type (see Eq. (3) in \citealt{Arguelles2018}) and it includes two phenomena. One is the Pauli exclusion principle (i.e. fermion degeneracy), causing a dense fermion core at the center of the halo that acts as an alternative to the SMBH scenario. On the other hand, it is the effect of escape of particles that leads to finite-sized halos. Once with such a most likely fermionic DF, the equation of state (E.o.S, $\rho(r), P(r)$) is built as corresponding momentum integrals of this DF, in order to obtain the DM profile by solving a coupled system of equilibrium equations for appropriate boundary conditions taken from galactic observables. The equilibrium equations of the RAR model are obtained by solving the Einstein (general relativistic) equations in spherical symmetry, sourced by a perfect fluid ansatz with the above ($\rho(r), P(r)$) E.o.S, and read (in dimensionless form):

\begin{equation}
    \begin{aligned}
        \frac{d\hat{M}}{d\hat{r}} &= 4\pi\hat{r}^2\hat{\rho}, \\
        \frac{d\nu}{d\hat{r}} &= \frac{2(\hat{M}+4\pi\hat{P}\hat{r}^3)}{\hat{r}^2(1-2\hat{M}/\hat{r})}, \\
        \frac{d\theta}{d\hat{r}} &= -\frac{1-\beta_0(\theta-\theta_0)}{\beta_0}\frac{\hat{M}+4\pi\hat{P}\hat{r}^3}{\hat{r}^2(1-2\hat{M}/\hat{r})}, \\
        \beta(\hat{r}) &= \beta_0\mathrm{e}^{\frac{\nu_0-\nu(\hat{r})}{2}}, \\
        W(\hat{r}) &= W_0 + \theta(\hat{r}) - \theta_0.
    \end{aligned}
    \label{eq:RAR1}
\end{equation}
The dimensionless quantities are: $\hat{r} = r/\chi$, $\hat{M} = GM/(c^2\chi)$, $\hat{\rho}=G\chi^2\rho/c^2$, $\hat{P} =G\chi^2P/c^4$, with $\chi = 2\pi^{3/2}(\hbar/(mc))(m_p/m)$ and $m_p = \sqrt{\hbar c/G}$ the Planck mass. The system of Eqs. (\ref{eq:RAR1}) constitute an initial value problem which, for fixed DM particle mass $m$, has to be solved for a given set of free parameters ($\beta_0$, $\theta_0$, $W_0$) defined at the center of the configuration.

The first two are the only relevant Einstein equations (mass and T.O.V equations), the third and fourth are obtained from the Tolman and Klein relations for a gas in thermodynamic equilibrium in GR, and the fifth corresponds to particle energy conservation along a geodesic (see \cite{Arguelles2018} and refs. therein for details). The density and pressure ($\rho(r), P(r)$) of the system are varying (non-analytic) functions of the radius (see Eqs. $(2),(3)$ in \citealt{2021MNRAS.502.4227A}). The non-linear system of (ordinary) coupled differential equations written above, implies an initial condition problem for the variables $M(0) = M_{0} = 0$, $\nu(0) = \nu_{0}$, $W(0) = W_{0}$,  $\theta(0) = \theta_{0}$ and  $\beta(0) = \beta_{0}$, where the subscript $0$ indicates they are evaluated at the center of the configuration. The function $M(r)$ is the enclosed mass at radius $r$, and $\nu(r)$ is the metric potential matching with the Schwarzchild potential at the boundary, ensuring this condition through the value of $\nu_{0}$ (see \cite{2021MNRAS.502.4227A} for details).

The functions $W(r)$, $\theta(r)$, and $\beta(r)$ are named as the cut-off, degeneracy, and temperature variables, respectively. For an equivalent setup of the fermionic equations and for the numerical scheme actually used to solve them in this paper, see respectively Sec. 2.2 and Ap. A in~\cite{Mestre2024}.

Such a model depends on four free parameters, the degeneracy parameter $\theta_{0}$, the cut-off parameter $W_{0}$, the temperature parameter $\beta_{0}$ and the DM particle mass $m$. This set of free parameters will provide a solution for the mass distribution, whose characteristics strongly depend upon the galaxy type (see e.g. \citealt{2023ApJ...945....1K} for an allowed window of RAR free parameters in agreement with disk galaxies from the SPARC data-set, and \citealt{Arguelles2018,Arguelles2019} for the Milky Way and other galaxy types respectively). Because of this, we will generate different RAR halos whose gravitational potential (together with the baryonic counterpart) will be used to run the spray algorithm, allowing us to study how the different profiles change the features of the stream.

We will use four different profiles, three of them of fermionic nature named as RAR 1, RAR 2, RAR 3, and a Burkert one~\citep{Burkert1995}. Their specifications are explained as follows.

\begin{figure}
    \centering
    \begin{tabular}{@{}c@{}}
        \includegraphics[width=\columnwidth]{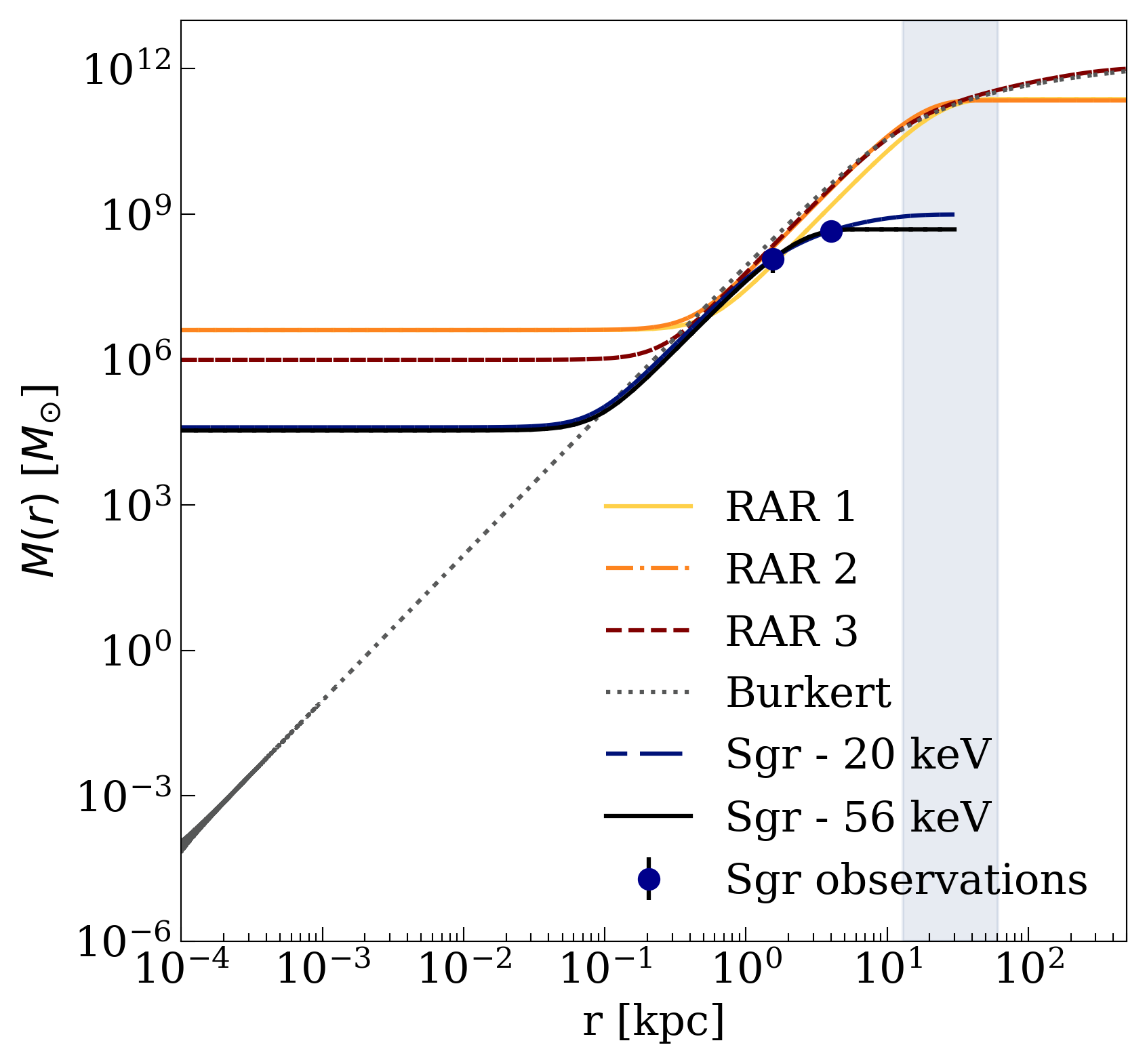}
    \end{tabular}
    \begin{tabular}{@{}c@{}}
        \includegraphics[width=\columnwidth]{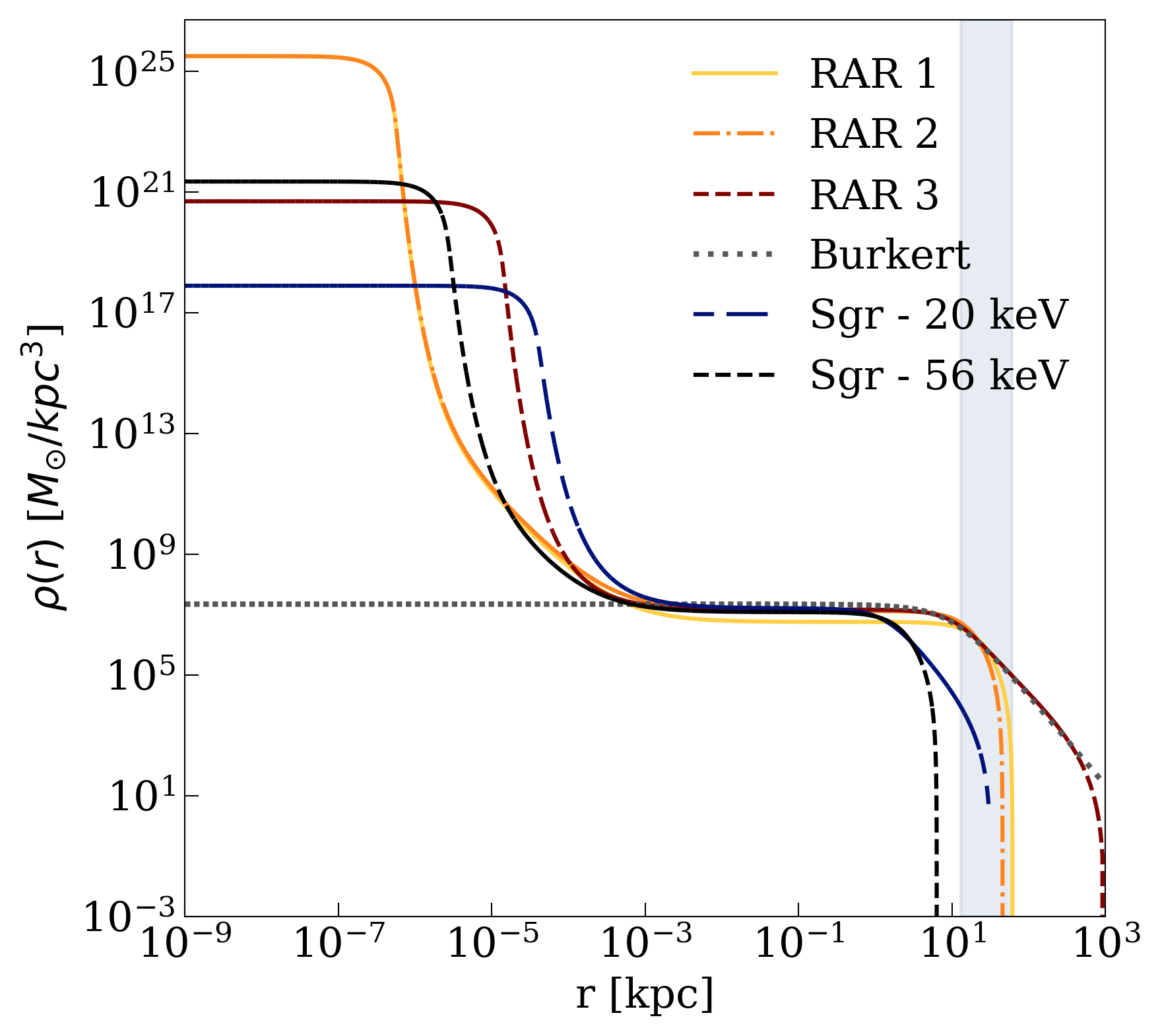}
    \end{tabular}
    \caption{Top panel: RAR enclosed mass profiles corresponding to the best-fit parameters given in Table \ref{tab:rar_best_fit_parameters} which satisfy the enclosed mass values in Table~\ref{tab:accumulated_mass} and \cref{sec:progenitorPot}, respectively for the MW and the Sgr dwarf. Additionally, we display the fitted Burkert profile for comparison. Bottom panel: profiles corresponding to the density of the same dark matter halos considered in the upper panel. In both panels, the shaded region indicates where the progenitor will move in its orbit around the Galaxy.}
    \label{fig:profiles}
\end{figure}

\begin{itemize}
    \item The DM model RAR 1 is defined according to the set of free parameters as found in \cite{BecerraVergara2020,2021MNRAS.505L..64B} for the Milky Way. For such a set of parameters, with $m=56$ keV, it was proved that the DM halo is of core--halo nature, where the dense and compact DM core can fit the orbits of the S-cluster stars without assuming a BH \citep{2021MNRAS.505L..64B}. At the same time, the outer halo develops a tail of polytropic trend such that the overall profile can reproduce the so-called Grand Rotation Curve of the Galaxy (ranging from pc scales up to tens of kpc scales according to the data provided in \cite{Sofue2013}). The motivation for choosing this specific fermionic model (originally aimed to fit the rotation curve of the Galaxy together with the S-cluster stars orbits) was just to check how it stands with respect to an independent tracer of the Galactic potential, such as the Sgr stream. The accumulated DM mass of this model fulfills the specific values at four different radii (central, inner, mid, and outer) as reported in Table \ref{tab:accumulated_mass}.

    \vspace{0.085in}

    \item The DM models RAR $2$ and RAR $3$ are asked to fulfill (within $2 \sigma$ dispersion) with selected accumulated total mass values as inferred in \cite{Gibbons2014} from the observed apocentric distances of the leading and trailing arms of the stream and their opening angle (see Fig.~13 and Table 3 of that work), and reported in Table \ref{tab:accumulated_mass}. These mass values are close to the mass constraints from~\cite{Vasiliev2021}: $M(r=50~\mathrm{kpc})=(3.85\pm0.1)\times10^{11} M_\odot$ and $M(r=100~\mathrm{kpc})=(5.7\pm0.3)\times10^{11} M_\odot$. RAR $2$ and RAR $3$ differ from the RAR $1$ case on outer halo scales as follows: the DM mass of the RAR $2$ profile at $12$ kpc is roughly twice as massive than the RAR $1$ at the same radius, while RAR $3$ is almost $70 \%$ more massive than RAR $1$ at the same scale (the mass of the fermion core is asked to be the same in all cases). Since the leading arm apocenter is most sensitive to the accumulated mass profile in the $[10-60]$ kpc radial window, then considerable differences in the behaviour of $M(r)$ in such a window are expected to lead to noticeable changes in the predictions of the shape of the corresponding part of the stream (see e.g. Figs. \ref{fig:dist_lambda} to \ref{fig:losvel_vs_L}). Another difference between RAR $2$ and RAR $3$ on halo scales is the accumulated mass at the outermost radii $80$ kpc, which in the latter case is roughly twice larger than in the former. Given the allowed physics of the fermionic model, this implies that the morphology of the outer halo tail of RAR $3$ develops a more extended power-law-like trend similar to Burkert (and different from polytropic). A thing like that is achieved for a lower particle mass than in the first two RAR cases. Finally, since the compacity of the dark matter core sensitively depends on the fermion mass, only the solutions of $m=56$ keV can provide enough compact cores to work as a good alternative to the BH in SgrA* as demonstrated in  \cite{BecerraVergara2020,2021MNRAS.505L..64B,2022MNRAS.511L..35A}.

    \vspace{0.085in}

    \item The Burkert profile is a two-free parameter dark matter density profile with a power law behavior $\propto r^{-3}$ for the halo tail, given by the equation:
    \begingroup
    \large
    \begin{equation}
        \rho_{\mathrm{Bur}}(r) = \frac{\rho_{0}}{\left(\frac{r}{r_{0}} + 1 \right) \left[\left(\frac{r}{r_{0}}\right)^{2} + 1 \right]},
        \label{Burkert}
    \end{equation}
    \endgroup
    and shown in Fig.~\ref{fig:profiles}. It is chosen in order to make a comparison between a typical model used in the literature obtained within $\Lambda$WDM cosmologies, with the (similar) power-law-like trend of the DM halo in the RAR $3$ case. The density parameter $\rho_{0}$ is the density at the origin of the distribution, i.e., $\rho_{0} = \rho(0)$. On the other hand, $r_{0}$ corresponds to a halo-scale radius where the density satisfies $\rho(r_{0}) = \rho_{0}/4$. The values of the best-fit parameters from this model are given in Table \ref{tab:rar_best_fit_parameters} in agreement with the specific accumulated masses of Table \ref{tab:accumulated_mass}. See Sec. \ref{sec:comparison} for a discussion about the similitude in the predictions of the stream between Burkert and the RAR 3 model.

\end{itemize}

To find the best values of the free parameters in the RAR and Burkert models, we used a {\it differential evolution} algorithm\footnote{We have used an implementation from the {\it SciPy} libray~\citep{2020SciPy-NMeth}, called \texttt{optimize.differential\_evolution} algorithm, with metaparameters given by \texttt{strategy}=\texttt{"best2bin"}.} to fit the predicted enclosed mass to the values in Table~\ref{tab:accumulated_mass}, adopting an allowed ``error'' in fitting such boundary mass values to the RAR models of $\sim 10\%$. The corresponding best-fit parameters ($\beta_0,\theta_0,W_0$) for the fermionic models, and ($\rho_{0}$, $r_{0}$) for the Burkert one are given in Table \ref{tab:rar_best_fit_parameters}, while the corresponding mass and density profiles are given in Fig.~\ref{fig:profiles}.

\begin{table*}
    \caption{Free parameter values.}
    \centering
    \begin{tabular}{ccccccc}
        \hline
        Parameter & RAR $1$ & RAR $2$ & RAR $3$ & Burkert & Sgr20 & Sgr56 \\
        \hline
        $mc^{2}$ [keV] & $56$ & $56$ & $20$ & $-$ & $20$ & $56$ \\
        $\theta_{0}$ & $37.766$ & $37.416$ & $34.417$ & $-$ & $25.919$ & $31.611$ \\
        $W_{0}$ & $66.341$ & $65.745$ & $68.835$ & $-$ & $52.236$ & $56.065$ \\
        $\beta_{0}$ & $1.198\cdot 10^{-5}$ & $1.208\cdot 10^{-5}$ & $1.274\cdot 10^{-7}$ & $-$ & $2.312\cdot 10^{-9}$ & $2.436\cdot 10^{-8}$ \\
        $\rho_{0}\ [M_{\odot}/\mathrm{kpc}^{3}]$ & $-$ & $-$ & $-$ & $2.197\cdot 10^{7}$ & $-$ & $-$ \\
        $r_{0}\ [\mathrm{kpc}]$ & $-$ & $-$ & $-$ & $10.087$ & $-$ & $-$ \\
        \hline
    \end{tabular}
    \tablefoot{Free parameters of the three RAR models and the Burkert profile which better predict the enclosed mass values given in Table \ref{tab:accumulated_mass}, as indicated in the text. Also, it is shown the best RAR free parameters which fit the enclosed Sgr mass according to the values given in Sec. \ref{sec:potential}. Sgr20 corresponds to the model with a dark matter particle mass of $20$ keV and Sgr56 to the one with a mass of $56$ keV.}
    \label{tab:rar_best_fit_parameters}
\end{table*}

\subsection{Potential of the progenitor}
\label{sec:progenitorPot}

It has been proven that the gravitational attraction of the satellite is needed to correctly reproduce the stellar stream ejected by Sgr. For example, \cite{Gibbons2014} used a spray method to generate a stellar stream and studied the same tidal stream with and without the self-gravity of the satellite. The resulting stream embedded in the gravitational potential of the host and the progenitor agreed with the results of the $N-$body simulations they made.

To model the progenitor of the stream, we considered that it is formed by two components, a baryonic component and a dark matter one. This choice is suggested in \cite{Vasiliev2020a}, where it is argued that a single baryonic component progenitor cannot match all the observational constraints. We adopted a Plummer sphere of mass $10^{8}\ M_{\odot}$ and scale length $b_{\mathrm{Sgr}} = 0.3$ kpc to model the baryons of this galaxy. Afterwards, we performed a best fit on a RAR dark matter halo constrained by two data points for the accumulated DM mass of the Sgr galaxy: $M(1.55~\mathrm{kpc})=(1.2\pm 0.6)\times 10^{8} M_{\odot}$ and $M(4~\mathrm{kpc})=(4.5\pm0.7)\times 10^{8} M_{\odot}$. The former constraint was taken from \cite{Walker2009} while the latter one is in agreement with the value estimated in \cite{Vasiliev2020a}: $M(5~\mathrm{kpc})=(4\pm1)\times 10^8 M_\odot$. Due to the facts that (i) our spray algorithm uses a constant progenitor potential, and that (ii)  we are interested in studying specially the last $3~\mathrm{Gyr}$ of evolution, we decided that it was more realistic to model the progenitor according to current observations instead of using estimated values at infall.

Of course, the particle mass for the halo of the progenitor is always chosen, for consistency, to be the same as the one involved in the dark matter of the given host (for each of the three RAR cases). In the first two cases ($m=56$ keV), the halo model for the progenitor is named as Sgr56, and Sgr20 is used for the RAR $3$ ($m=20$ keV) and Burkert models. The enclosed mass values and density profiles of the Sgr models are shown in \cref{fig:profiles}.

\subsection{Orbit of the progenitor in the Galactic potential}
\label{sec:progenitorOrbit}

Since we will use a spray algorithm to model the stellar stream generated by Sgr in its orbit all around the host, we need first to compute the path of the progenitor backward in time. This is done before evolving the corresponding trajectory forward in time while ejecting stars from the progenitor. This way, the orbit of the ejected stars can be integrated under the combined gravitational potential of the host and the satellite.

To integrate the progenitor orbit, a set of initial conditions is needed to give to the integrator. These are the position and velocity of the progenitor, which were taken from \cite{Gibbons2014}, except for the heliocentric line of sight velocity, which was taken from \cite{Vasiliev2020a}. These collected observables are detailed in Table \ref{tab:ci}.

Since the Galactocentric distance of the progenitor is not very well constrained, we let it vary through the different models to get orbits with approximately the same apocenter, chosen to agree with the values obtained in \cite{Law2005}. In either case, they imply small differences in the pericentric distance, as can be seen in Table \ref{tab:integration_parameters}. Also, the time intervals for backward integration were taken slightly different among the models in order to obtain similar initial positions for the spray algorithm. All these considerations, together with the predicted circular velocity at the Galactocentric solar distance $R_{\odot}$ for each MW model, are summarized in Table \ref{tab:integration_parameters}.

\begin{table}
    \caption{Galactic coordinates of the Sagittarius dSph satellite.}
    \centering
    \begin{tabular}{cc}
        \hline
        Property & Value \\
        \hline
        Galactic latitude [deg] & $-14.1669$ \\
        Galactic longitude [deg] & $5.5689$ \\
        Heliocentric distance [kpc] & $22.0 - 28.4$ \\
        Heliocentric LOS velocity [km/s] & $142.0$ \\
        Proper motion $\mu_{b}$ [mas/yr] & $1.97\pm 0.3$ \\
        Proper motion $\mu_{l}\mathrm{cos}b$ [mas/yr] & $-2.44\pm 0.3$ \\
        \hline
    \end{tabular}
    \tablefoot{All the values above were taken from \cite{Gibbons2014}, except the heliocentric line of sight velocity, which was taken from \cite{Vasiliev2020a}.}
    \label{tab:ci}
\end{table}

\begin{table}
    \caption{Orbital parameters of the Sagittarius dSph.}
    \centering
    \begin{tabular}{cccccc}
        \hline
        Model & $d$ & $d_{a}$ & $d_{p}$ & $t$ & $v_c(R_\odot)$ \\
        \hline
        RAR 1 & $24.57$ & $60.21$ & $13.70$ & $-3.0$ & 199.4 \\
        RAR 2 & $25.64$ & $60.17$ & $14.35$ & $-3.22$ & 215.0 \\
        RAR 3 & $25.88$ & $60.22$ & $14.87$ & $-2.7$ & 213.4\\
        Burkert & $25.3$ & $60.14$ & $14.27$ & $-2.75$ & 214.5 \\
        \hline
    \end{tabular}
    \tablefoot{The values in the table correspond to the orbit integrated backward in time under different gravitational potentials. $d$ stands for the initial distance to the progenitor, $d_{a}$ for apocentric distance, $d_{p}$ for pericentric distance, $t$ is the backwards integration time, and $v_c(R_\odot)$ is the circular velocity at $R_\odot$. Distances are in kpc, the time in Gyr, and the circular velocity in km/s.}
    \label{tab:integration_parameters}
\end{table}

Before integrating the orbit, the initial conditions are transformed to the Galactocentric reference frame, using the parameters detailed in Sec. \ref{sec:observations}. Summing up, we will integrate four progenitor orbits, all of them sharing the same baryonic potential but each one with a different dark matter halo.

\begin{figure}
    \centering
    \includegraphics[width=\columnwidth]{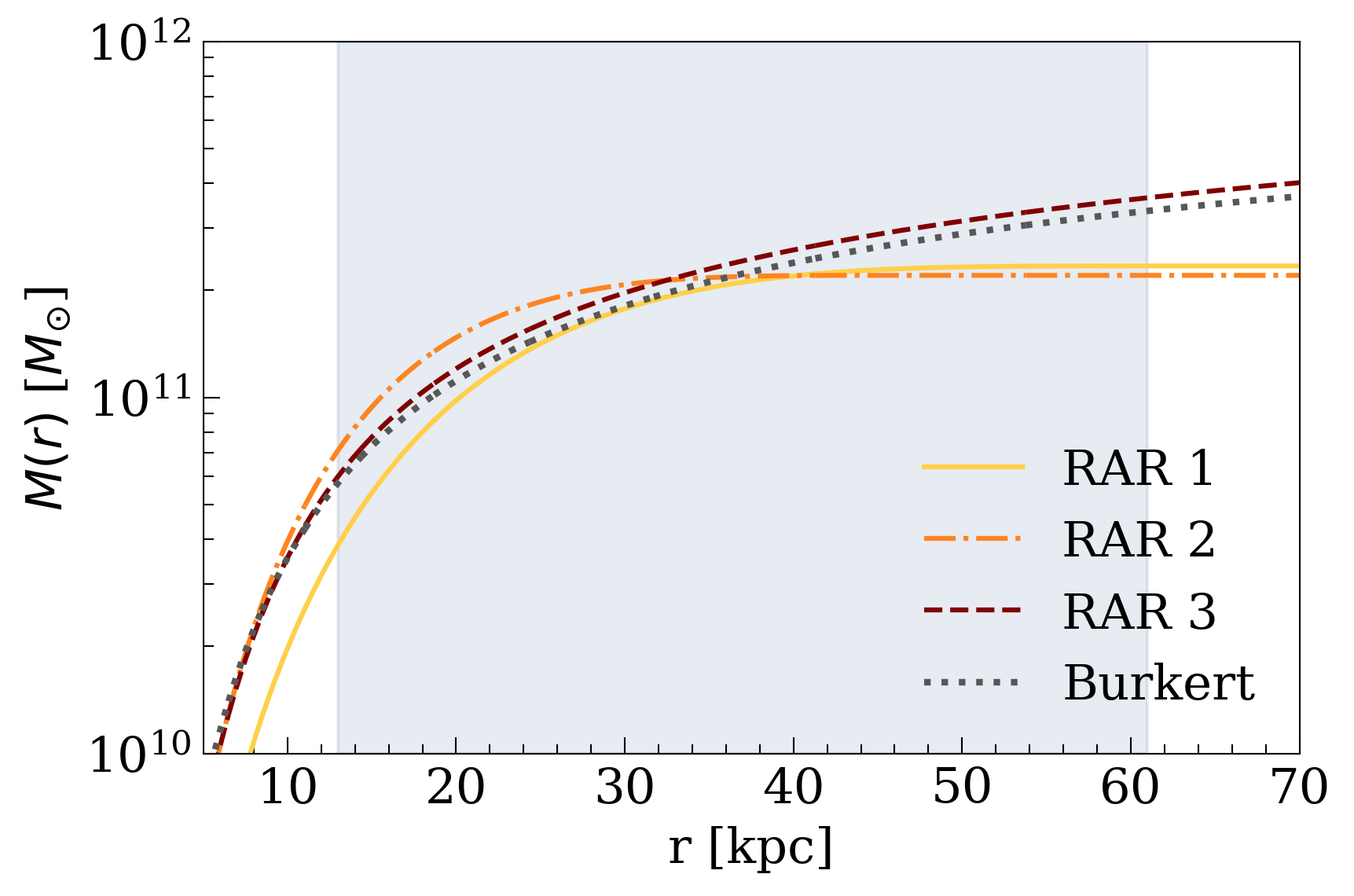}
    \caption{Enclosed masses for the four dark halo models studied, but in the Galactocentric scales where the orbit of the satellite resides. The shaded region denotes the limits of this orbit.}
    \label{fig:mass_profiles_orbit_region}
\end{figure}

\begin{figure}
    \centering
    \includegraphics[width=\columnwidth]{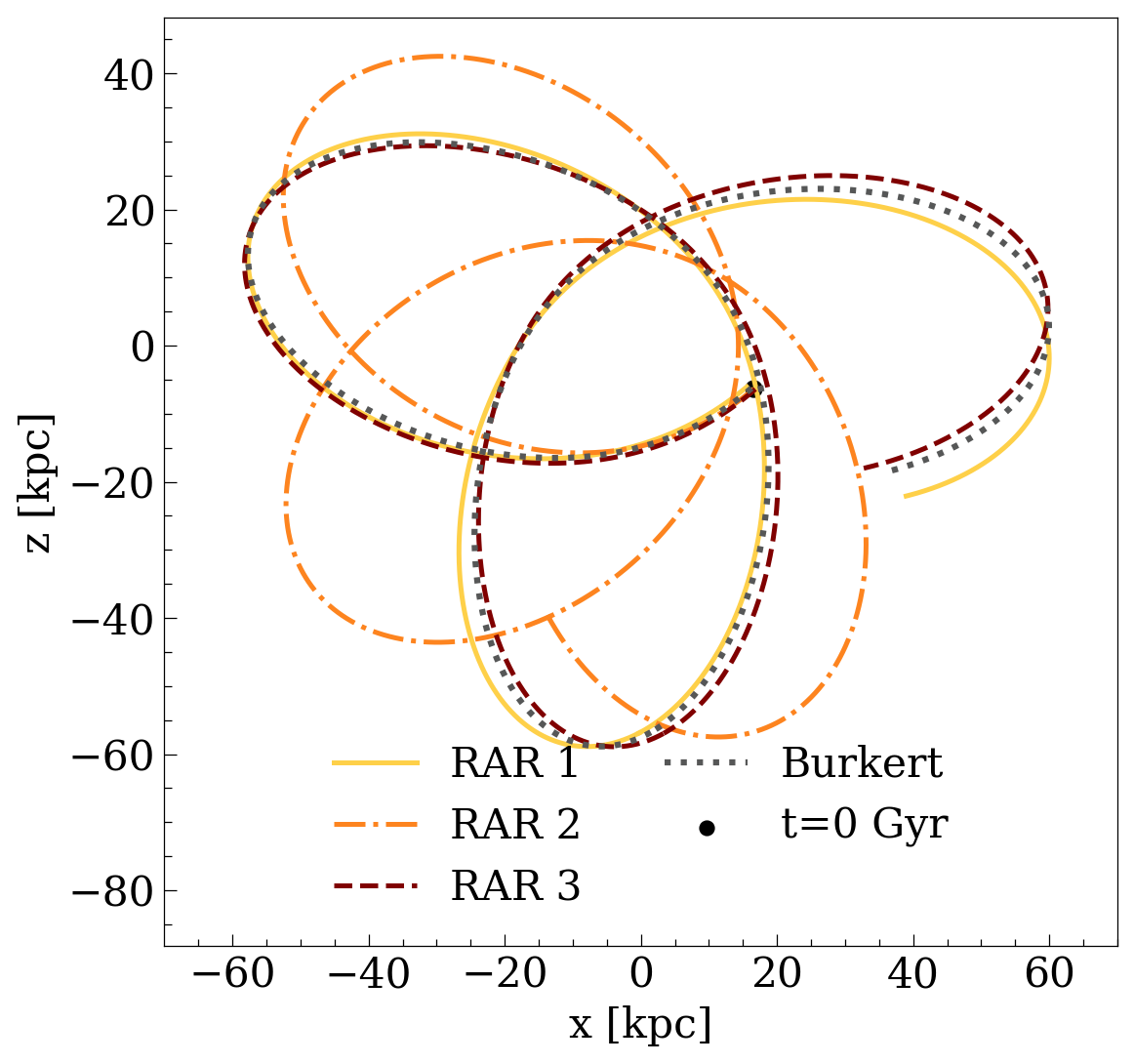}
    \caption{Orbits of the Sgr dSph integrated backward in time projected in the $x-z$ plane of the Galactocentric system. The orbits corresponding to the four models under consideration are shown. The black dot denotes the region where the backward integrations start.}
    \label{fig:orbit_xz}
\end{figure}

The integrated orbits of Sgr projected in the $x-z$ plane for the four different potentials are shown in Fig.~\ref{fig:orbit_xz}. Despite little differences in the orbits, they share the same general features, which are the number of pericentric passages, the area covered by each petal, the pericentric and apocentric distances, and (approximately) the time interval they took to complete the same number of petals. The small differences between the orbits arise because the gravitational field, under the assumption of spherical symmetry satisfied by the RAR and Burkert models, depends on the accumulated mass. This is better seen in Fig.~\ref{fig:mass_profiles_orbit_region}, where the shaded region marks the limits of the orbits. Inside this region, the four dark matter components have different accumulated masses, generating different gravitational fields along the radial window of interest. Notice that the RAR 2 model, which is the most massive in the 20-30 kpc range, produces the orbit with the fastest precession rate.

\section{Stream generation}
\label{sec:stream_generation}

A spray algorithm is a procedure where the initial conditions of stars belonging to the progenitor are generated to be integrated with the combined gravitational potential of the progenitor and the host galaxy. Since each initial condition generated represents a star, the number of initial conditions will define the number of stars that compose the stream. We will apply the spray algorithm as defined in \cite{Gibbons2014} but using a different prescription for the tidal radius ($r_t$).

The spray procedure provides the initial position and velocity of each star ejected from the progenitor. According to \cite{Gibbons2014}, the initial position of the stars at time $t$ corresponds approximately with the Lagrange points $L_{1}$ and $L_{2}$. We locate both points, following the standard procedure, at a $r_t$ distance from the center of the satellite, so the $L_{1}$ point will be located a Galactocentric radius $r - r_{t}$ while the $L_{2}$ point will be located at $r + r_{t}$.

Various works argue for an appropriate description of the tidal radius. We will follow the recipe from \cite{Gajda2016}, where it is derived an expression of the tidal radius for a satellite orbiting a host galaxy allowing for a rotating progenitor. This recipe has the form:

\begin{equation}
    r_{t} = r\left(\frac{[m(r_{t})/M(r)]\lambda(r)}{2\Omega_{s}/\Omega - 1 + [2 - p(r)]\lambda(r)}\right)^{1/3}.
    \label{eq:tidal_radius}
\end{equation}

The denominator of the above expression must be greater than zero; otherwise, the tidal radius does not exist. In such a case, the acceleration points towards the satellite and the tidal radius is interpreted as being infinite, which is precluded because \cite{Gajda2016} assumed that $r_{t} \ll r$. The quantities $\lambda(r)$ and $p(r)$ are $\lambda(r) = \Omega_{\mathrm{circ}}^{2}(r)/\Omega^{2}$ and $p(r) = d\ \mathrm{ln}M(r)/d\ \mathrm{ln}r$, respectively. $\Omega$ is the instantaneous angular velocity of the satellite, $\Omega_{\mathrm{circ}}(r)$ is the angular velocity of a circular orbit at radius $r$ and $\Omega_{s}$ is the angular velocity of a given star respect to the satellite. Assuming that the stars to be ejected are located at the instantaneous Lagrange points, we set $\Omega_{s}=\Omega$. Besides, we approximate $m(r_{t})$ with  the total mass of the satellite, $m_{\mathrm{sat}}$.

\begin{figure}
    \centering
    \includegraphics[width=\columnwidth]{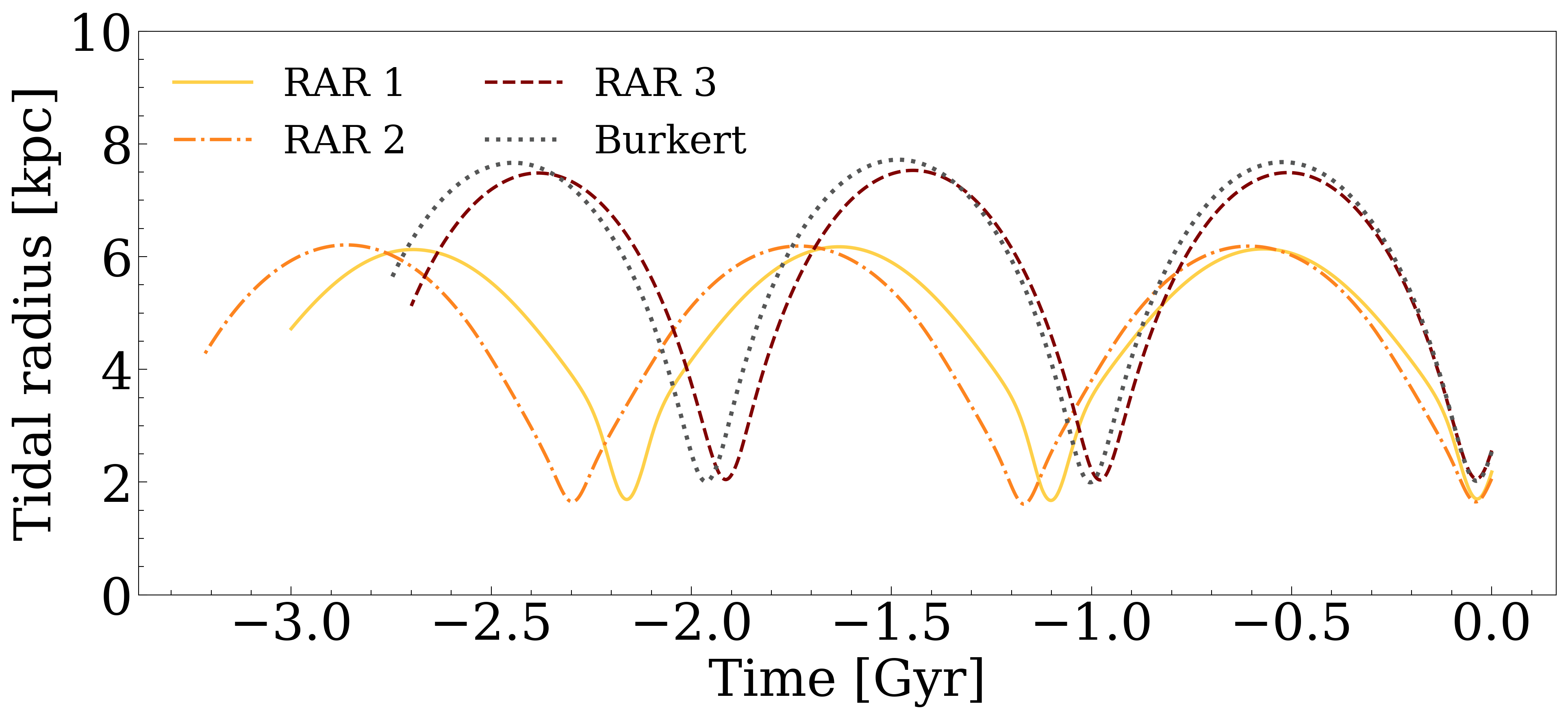}
    \caption{Tidal radius evolution for each model of gravitational potential considered in this work. The time intervals where each one is computed differ according to the parameters given in Table \ref{tab:integration_parameters}.}
    \label{fig:tidal_radius}
\end{figure}

Before proceeding, it is necessary to remark on an important assumption regarding the tidal radius. The model of \cite{Gajda2016} was derived under spherical symmetry, and we are modeling the Milky Way disk as composed of two baryonic axisymmetric components. Hence, to avoid such inconsistency, we will compute the logarithmic derivative of $M(r)$ in the formula for $r_t$ considering only the DM (spherical) component. This assumption is, in part, motivated by the estimated value of the radius of the baryonic disk of the Milky Way. It has been argued in different papers \citep{Sale2010, Minniti2011} that the stellar density of stars in the Galaxy has a drastic drop-off at a Galactocentric distance $R \sim 13$ kpc, which is the same order as our pericentric radius. The temporal evolution of the tidal radius of each gravitational model of Table~\ref{tab:rar_best_fit_parameters} can be seen in Fig.~\ref{fig:tidal_radius}.

The value of the tidal radius varies along the given progenitor's orbit due to the variation in time of the different quantities on which the tidal radius depends, i.e., the instantaneous orbital angular velocity of the satellite and the Galactocentric distance. The condition $1 + [2 - p(r)]\lambda(r) > 0$ was checked in each time step where the tidal radius was computed.

Regarding the velocities with which the stars will be ejected, they will be taken randomly from a multivariate Gaussian distribution whose mean is the Galactocentric velocity of the satellite at that time, and the dispersion is the velocity dispersion tensor of the progenitor. To model the latter, we will assume it is diagonal and isotropic, i.e., all its components are the same, $\sigma$.

The Sgr dwarf spheroidal is far from dynamical equilibrium due to the strong tidal forces acting on it, and it has been a long time orbiting the Milky Way and being disrupted. So, when we mention the velocity dispersion of the Sgr galaxy, we are referring to the progenitor of the stream, which is the bound central region of the whole stellar structure. The value of $\sigma$ we will adopt is $\sigma = 11.4$ km/s as taken from \cite{Ibata1997a}.

We will generate four stellar streams, one for each Milky Way's gravitational potential. In order to do so, we will eject $\sim 10^{5}$ stars, one half from $L_{1}$ and the other half from $L_{2}$. The ejection process will be carried out at the same time the progenitor evolves in its orbit to the future, finishing its trajectory at the black dot point of Fig.~\ref{fig:orbit_xz}. Each ejection is represented by an initial position and velocity, following the recipe previously introduced. Once the initial condition is created, it is used as input for an integrator whose function is to compute the trajectory of such a star under host and satellite potentials. The integration will occur from the ejection time until the present ($t = 0$ Gyr). Since we are not considering the force exerted by the individual stars, the integration of the orbit of each of them is completely independent of the others, thus enabling the parallelization of the process.

\section{Comparing the model with data and discussion}\label{sec:comparison}

\begin{figure*}
    \centering
    \includegraphics[width=0.495\hsize]{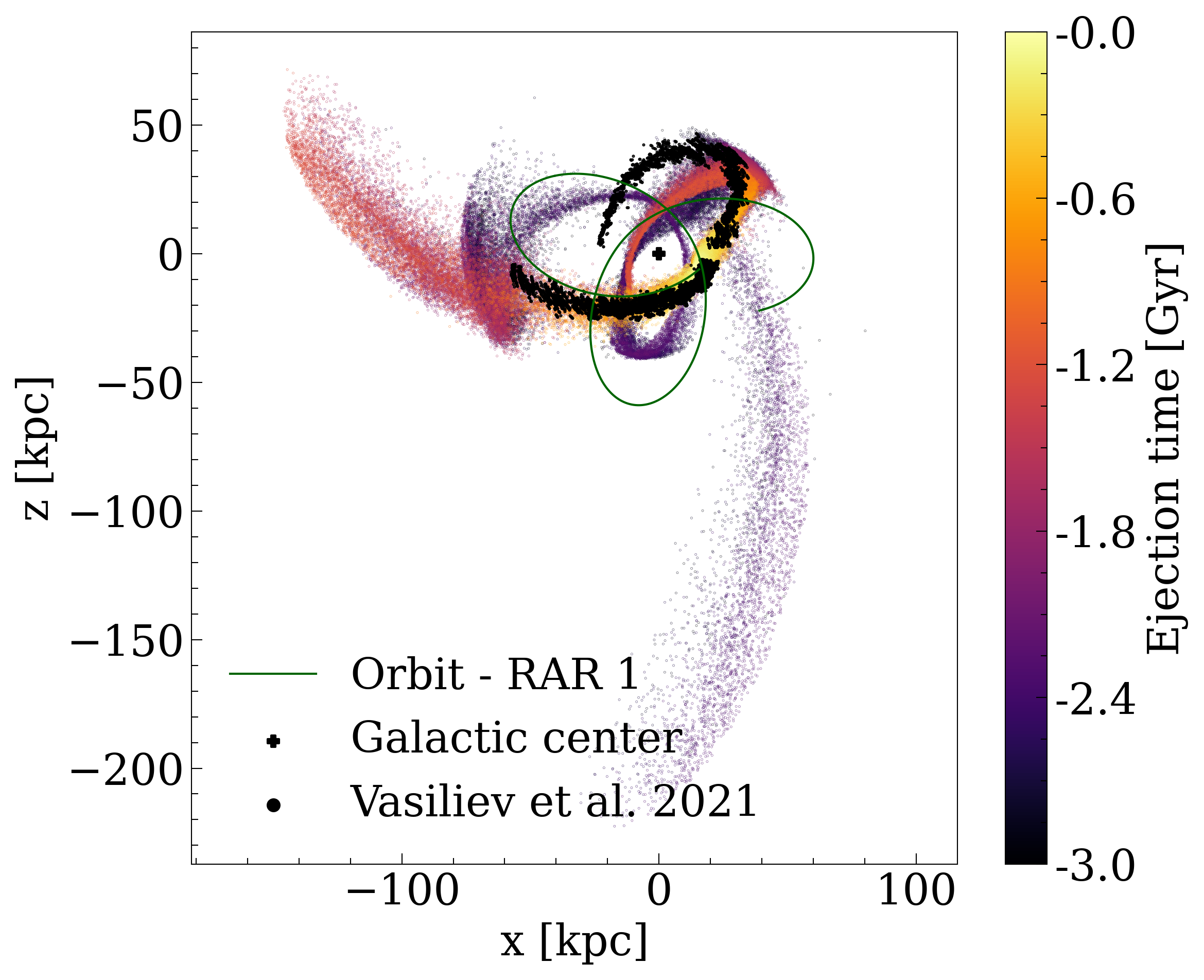}
    \hfill
    \includegraphics[width=0.495\hsize]{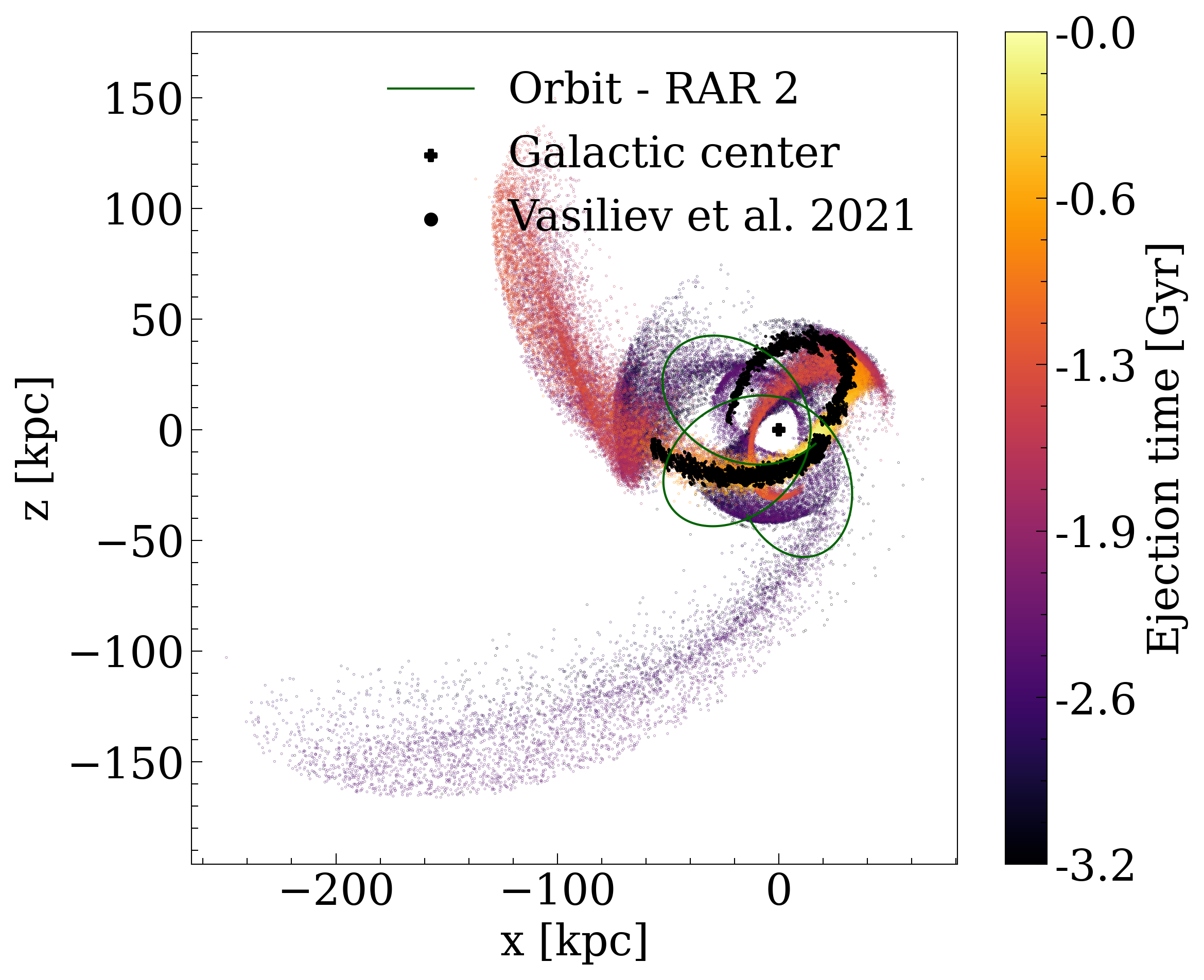}
    \vskip\baselineskip
    \includegraphics[width=0.495\hsize]{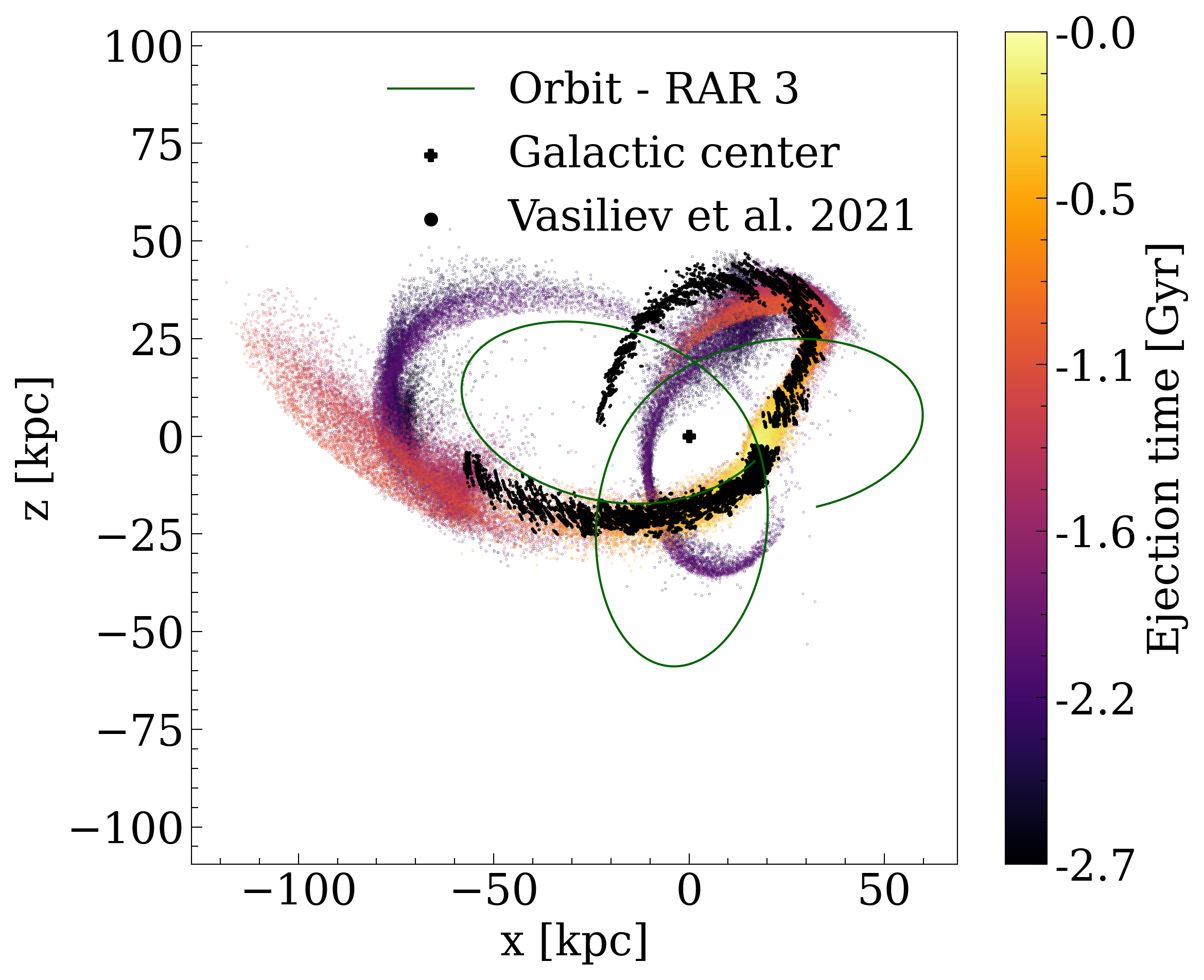}
    \hfill
    \includegraphics[width=0.495\hsize]{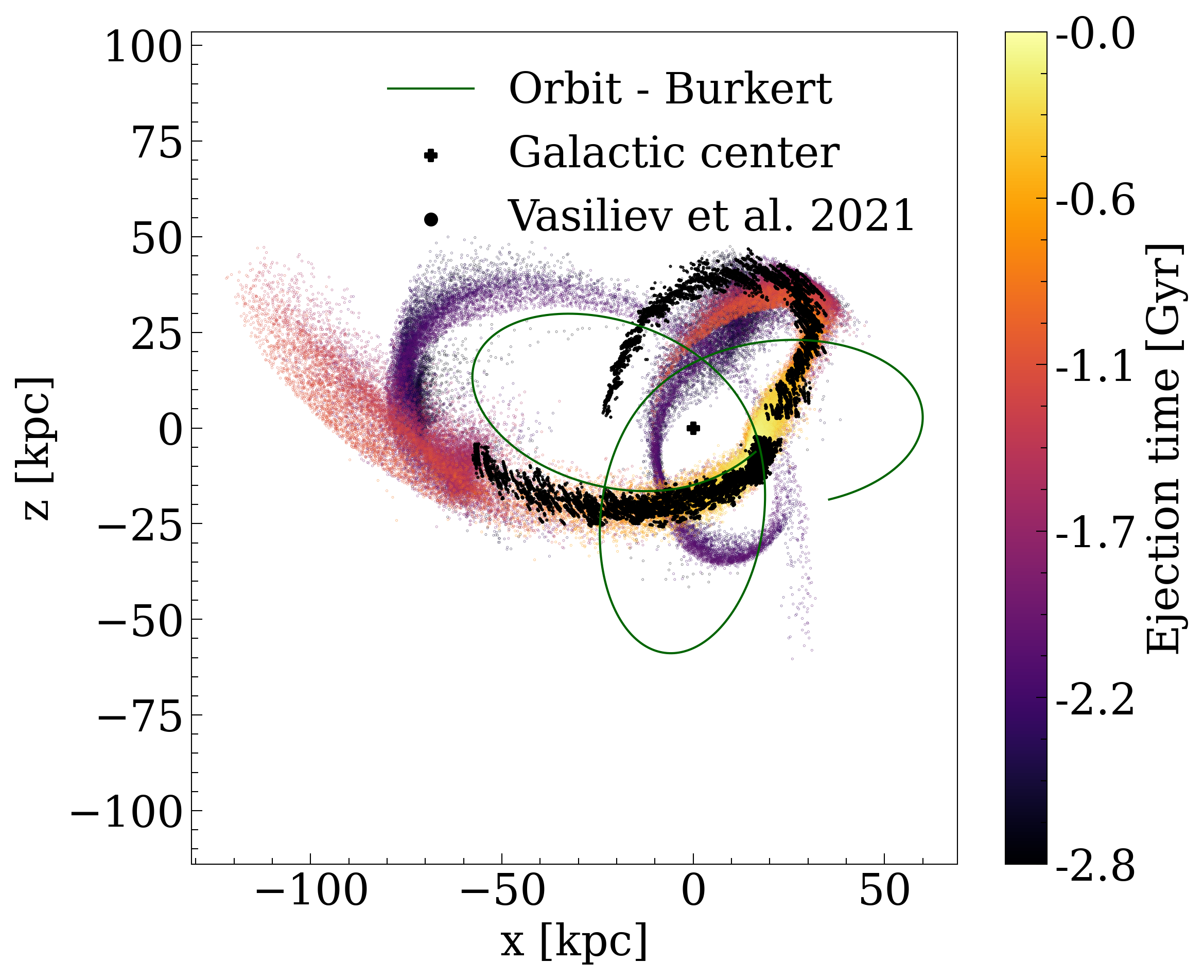}
    \caption{Streams of stars projected in the $x-z$ plane at present, color coded by the ejection time. The black dots represent the $x-z$ coordinates of the \cite{Vasiliev2021} stars. The green solid lines indicate the orbit of the satellite in its corresponding gravitational potential.}
    \label{fig:xz_time_of_ejection}
\end{figure*}

\begin{figure*}
    \centering
    \includegraphics[width=0.495\textwidth]{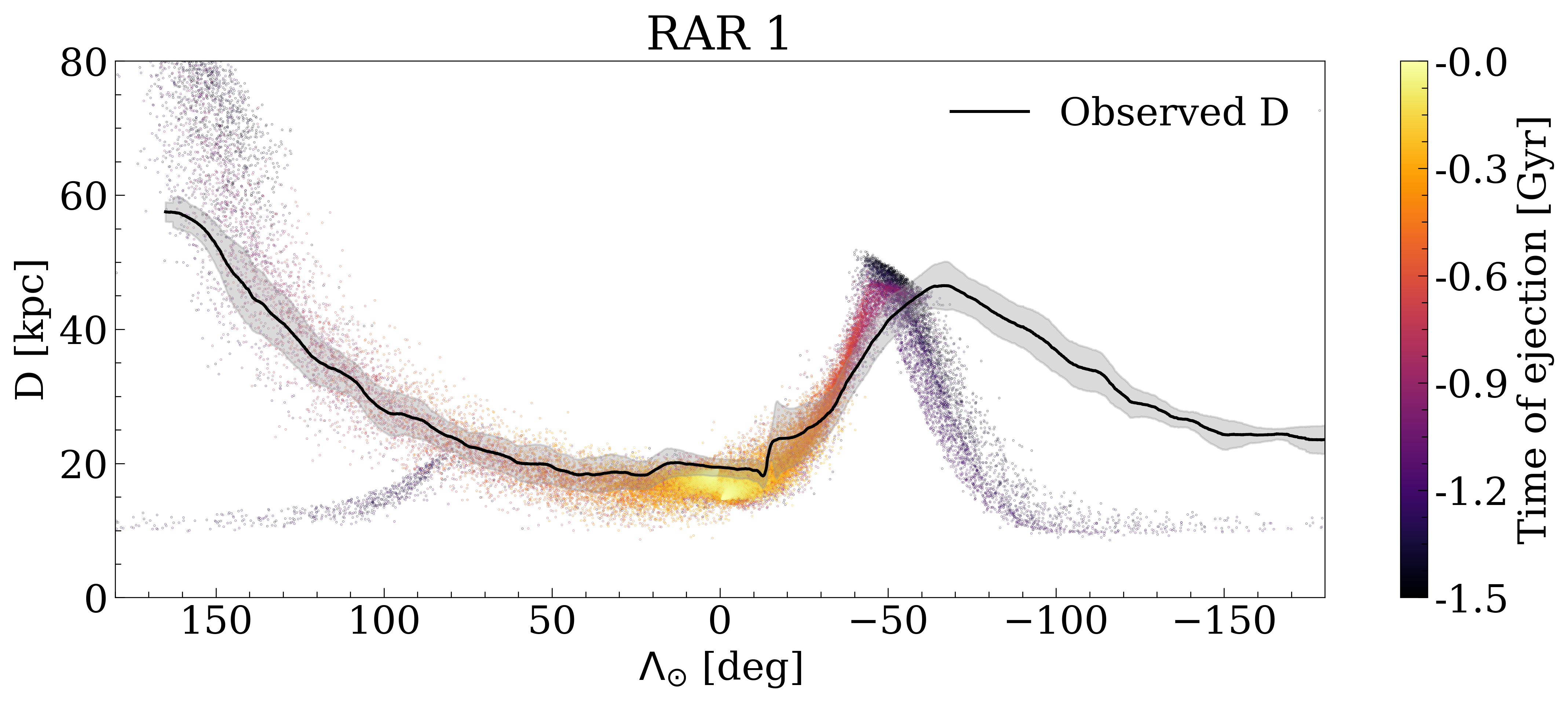}
    \hfill
    \includegraphics[width=0.495\textwidth]{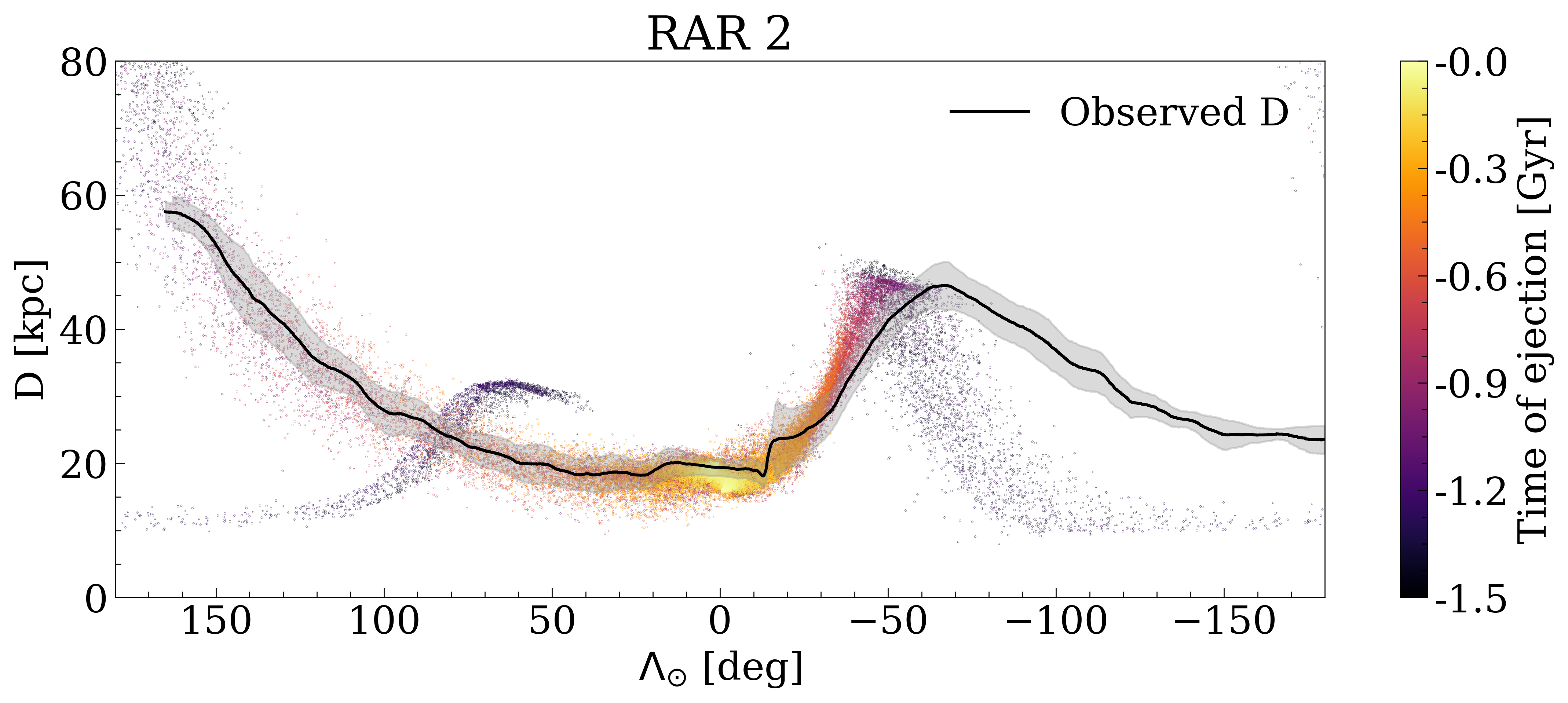}
    \vskip\baselineskip
    \includegraphics[width=0.495\textwidth]{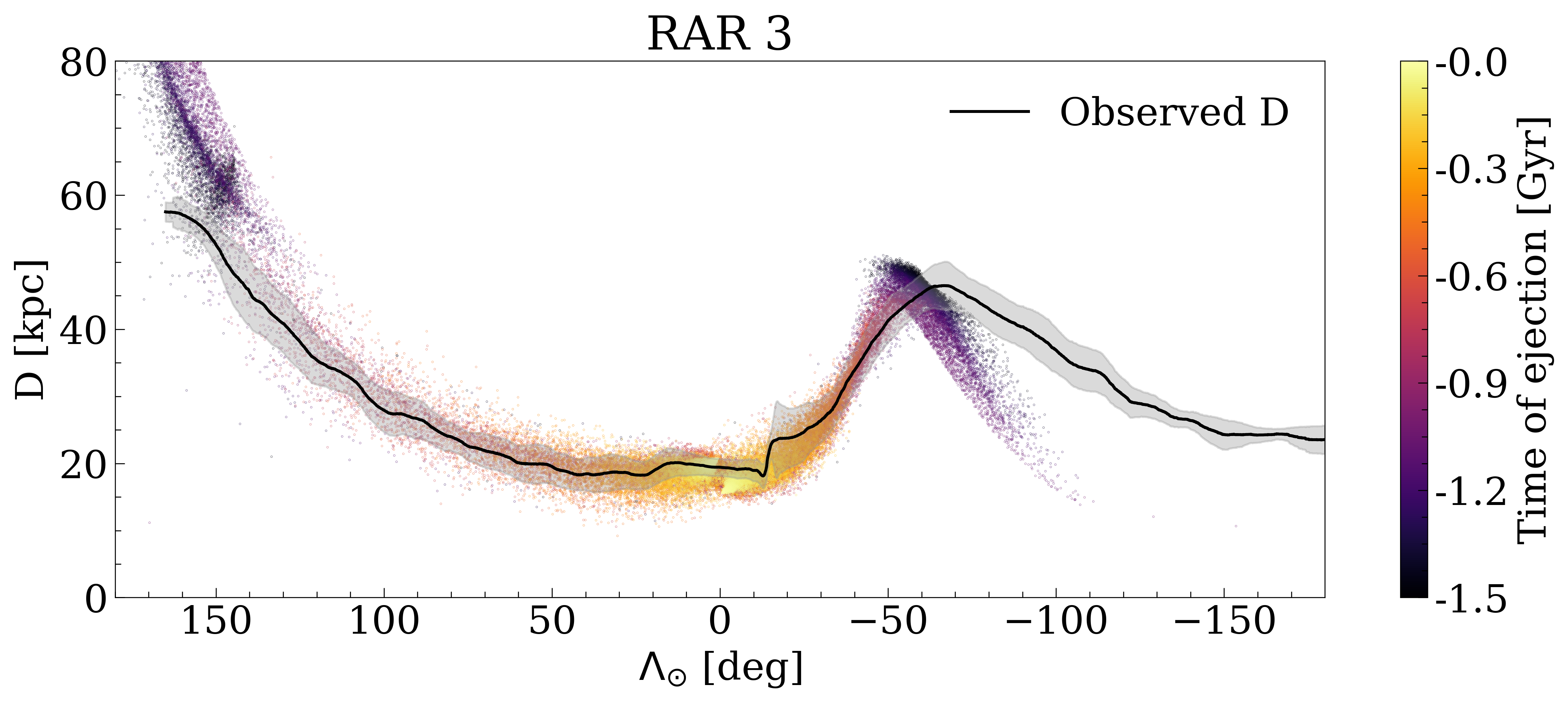}
    \hfill
    \includegraphics[width=0.495\textwidth]{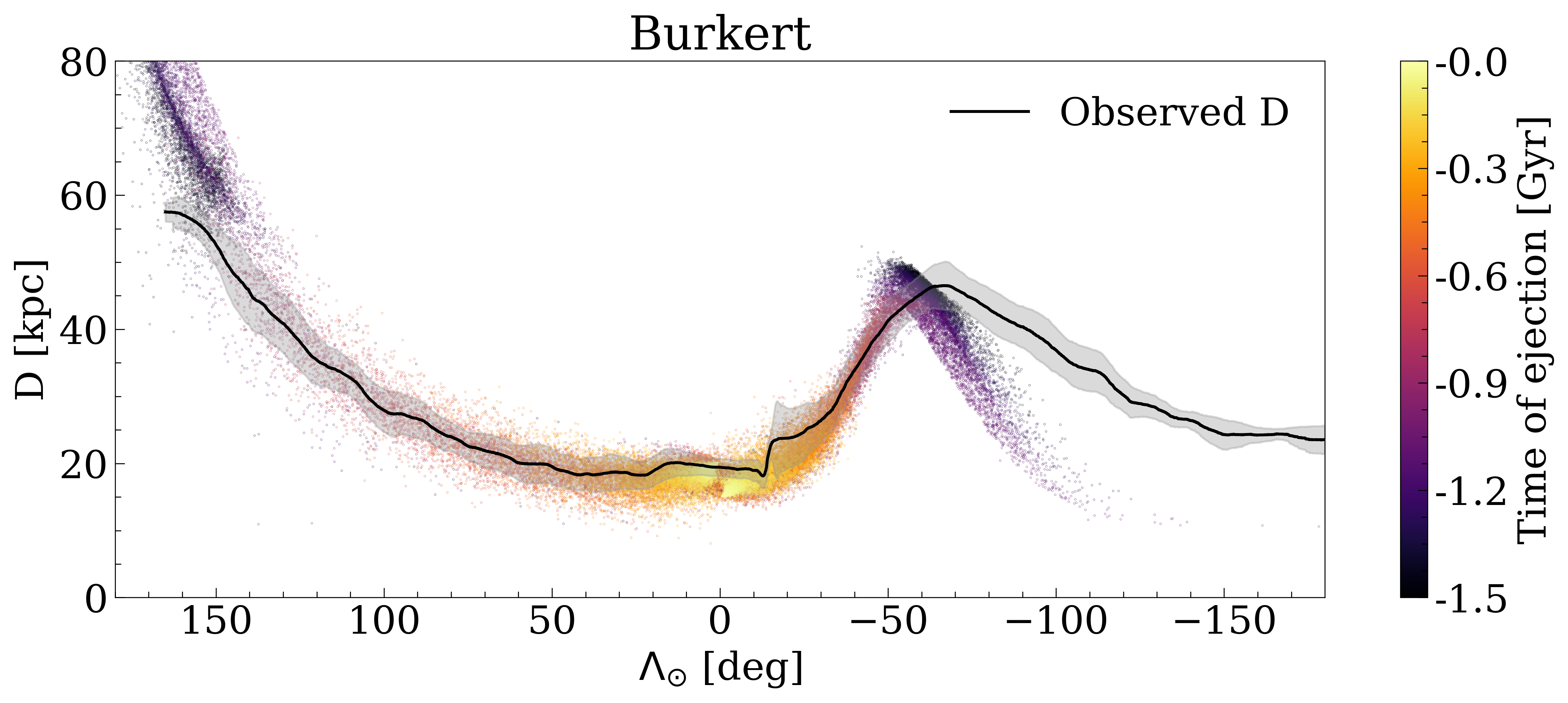}
    \caption{Streams of stars in the $D - \Lambda_{\odot}$ space at present color coded by the ejection time. The black solid curve represents the distance trend of the \cite{Vasiliev2021} stars, while the shaded region is the distance error, both quantities being moving means as a function of the longitude $\Lambda_{\odot}$. The distances are referred to the Galactocentric reference frame.}
    \label{fig:dist_lambda}
\end{figure*}

The final position of all the stars in the Galactocentric $x-z$ plane of the generated stream is shown in Fig.~\ref{fig:xz_time_of_ejection}, color-coded by the ejection time, starting $\sim 3$ Gyr in the past. The black dots and the solid green line represent the \cite{Vasiliev2021} stars and the past orbit of the progenitor. It can be seen that the different models produce different stream configurations of their tails (e.g. length, width, orientation), though they all agree on the fact of developing a clear bifurcation in their corresponding trailing tail. Interestingly, the RAR 2 stream configuration resembles the output of the N-body simulation shown in Fig.~2 of~\cite{Gibbons2014} with regards to the orientation of the trailing tail and the significant presence of the second wrap of their corresponding leading tails. In order to avoid the arising of the complementary and more feeble sinusoidal structure of the dynamically older part of the stream, where there is scarce data, in the rest of the plots we will display only mock stars stripped at times later than $t=-1.5$ Gyr. Transforming the four mock streams to the reference system of I$+$2020, in the configuration space we display the Galactocentric distance of the stars, $D$, as a function of $\Lambda_{\odot}$, as it is shown in \cref{fig:dist_lambda}.

With respect to the distribution of ejected stars in the $D - \Lambda_{\odot}$ space, it is clear from Fig. \ref{fig:dist_lambda} that neither of the four DM models can correctly reproduce the shape of the leading tail. All of them generate an arm that lies closer to the Galactic center than the observations. This behavior can also be seen in Fig. \ref{fig:xz_time_of_ejection}, where the leading tail reaches its maximum extent and then turns back instead of following the observed trend. Despite this discrepancy, it is worth noting that the RAR 3 and Burkert models (both power-law-like) achieve a slightly better agreement with the observed distance trend, making the mock stream fall less sharply with increasing negative longitude. Regarding the trailing arm, the four models agree very well with the observations, and among them, there are no remarkable differences in the predictions, as is the case for the leading tail.

\begin{figure*}
    \centering
        \includegraphics[width=\columnwidth]{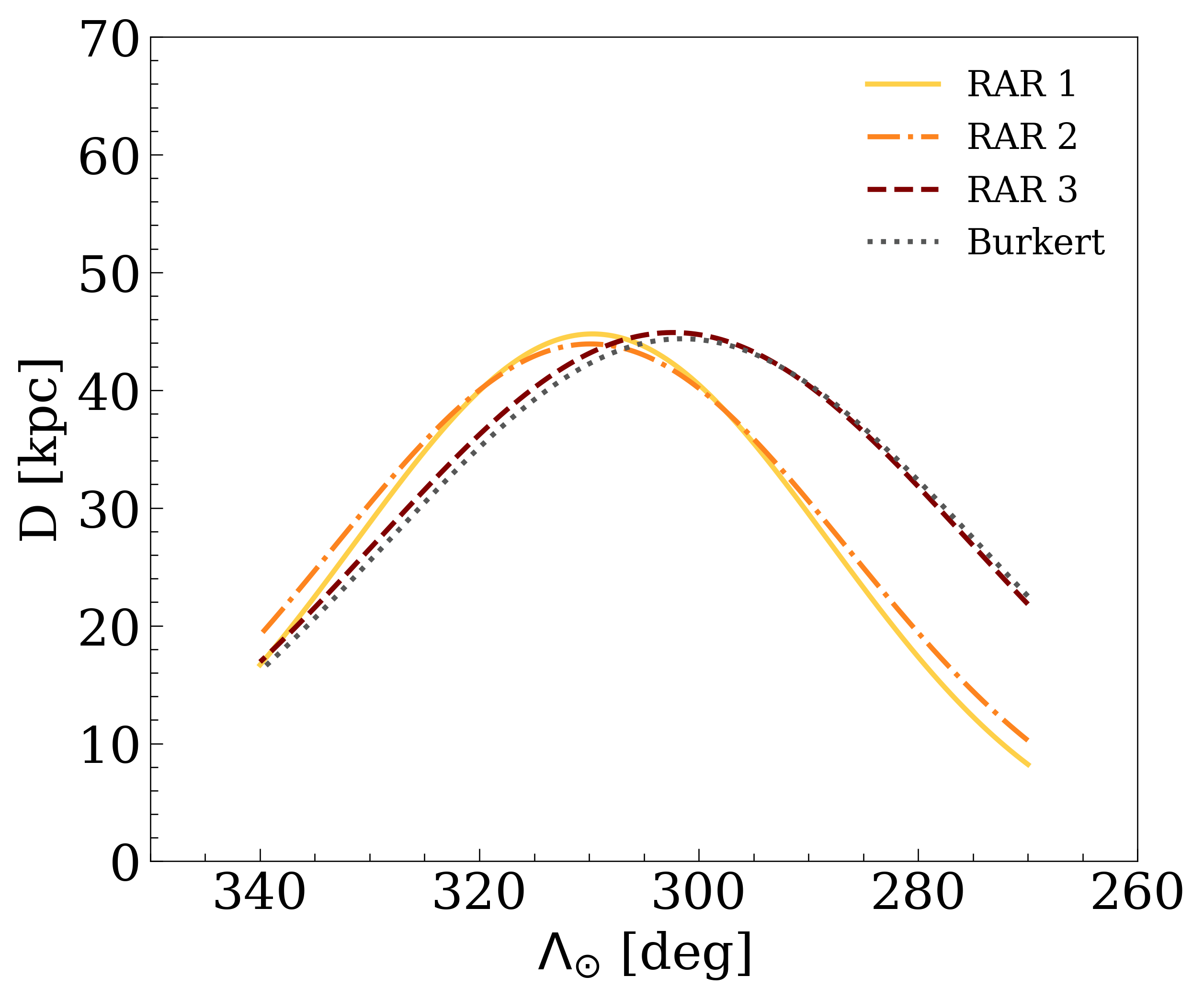}
    \hfill
        \includegraphics[width=\columnwidth]{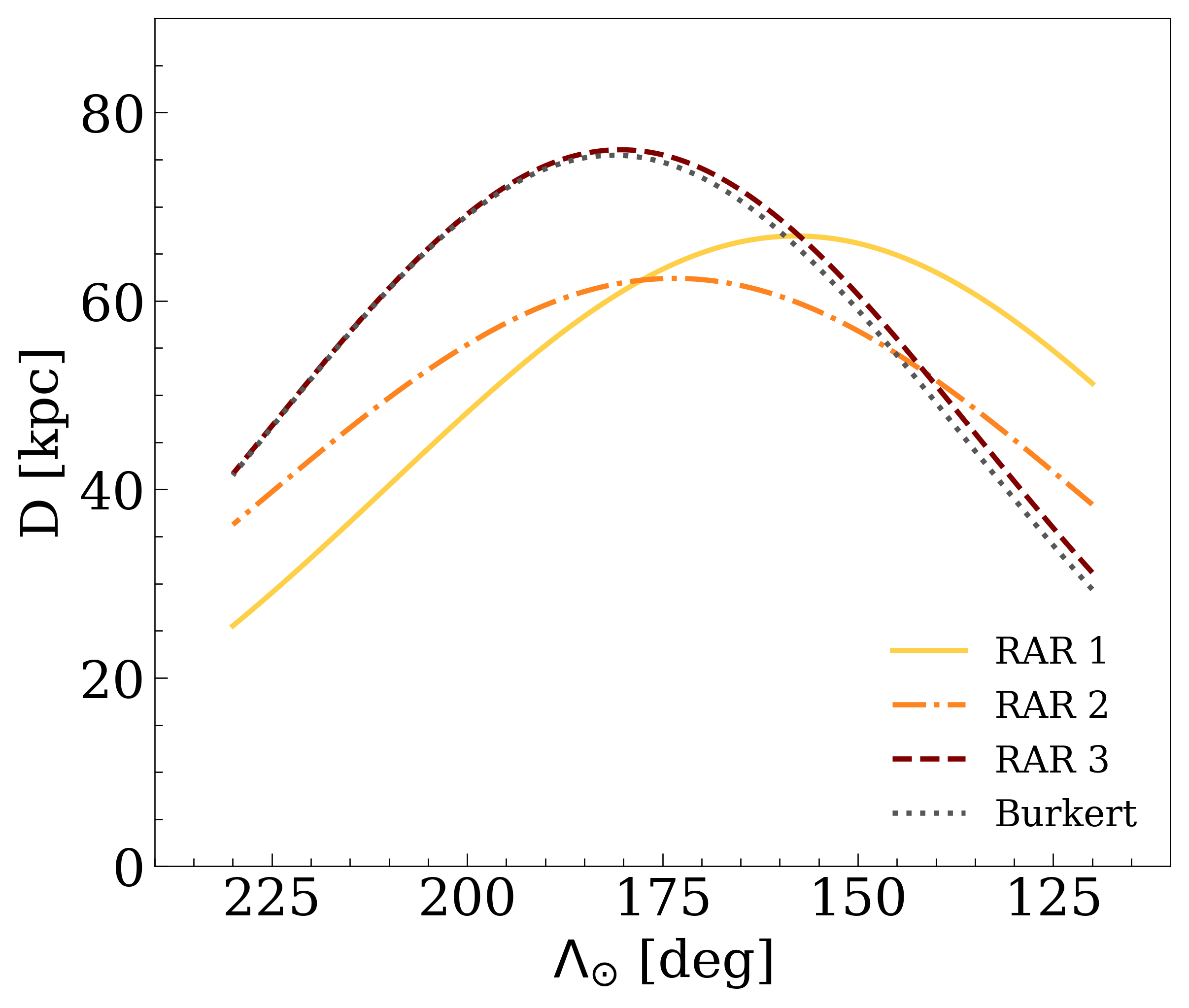}
    \caption{Analytic functions tracing the Galactocentric distance of both main tails of the Sgr stellar stream, the leading one (left plot) and the trailing one (right plot). Every curve is a Gaussian function, whose maximum represents the value of its corresponding apocentric distance. The location of the maximum is the longitude $\Lambda_{\odot}$ of the corresponding apocenter, used to compute the heliocentric angular aperture. There is one curve for each DM model considered in the work.}
    \label{fig:leading_and_trailing_apocenter_shapes}
\end{figure*}

Looking more deeply at Fig.~\ref{fig:dist_lambda}, it can be seen that the leading arm centered at $\Lambda_{\odot} \sim -50\degree$ has a `peak' shape. This phenomenon defines a maximum distance of the leading arm stars from the Galactic center. The same behavior occurs with the trailing arm. These two maximum distances define the leading and trailing (Galactocentric) apocentric distances, respectively. Thus, we have fit a Gaussian function to each distribution of stars composing the arms. To properly delineate the shape of the leading tail, we have considered stars ejected at $t \gtrsim -2$ Gyr. In the case of the trailing arm, we have taken those stars with an ejection time $t \lesssim -2$ Gyr. The corresponding Gaussian functions are plotted in Fig.~\ref{fig:leading_and_trailing_apocenter_shapes}, where there is one curve for each DM model considered. The apocentric distances associated with each model and each arm are given in Table \ref{tab:apocenters}, along with the heliocentric angular aperture between said apocenters.

Comparing the results of Table \ref{tab:apocenters} with the ones reported in \cite{Belokurov2014} and \cite{Hernitschek2017} it can be seen that our models underestimate the Galactocentric trailing apocentric distance (e.g. in $\sim 25\%$ for RAR 3), while overestimate the observed value of the heliocentric angular aperture (e.g. in $\sim 20\%$ for RAR 3). On the other hand, there is a much better agreement between our predicted leading apocentric distances and the ones reported in \cite{Belokurov2014} and \cite{Hernitschek2017}.

\begin{figure}
    \centering
    \includegraphics[width=\columnwidth]{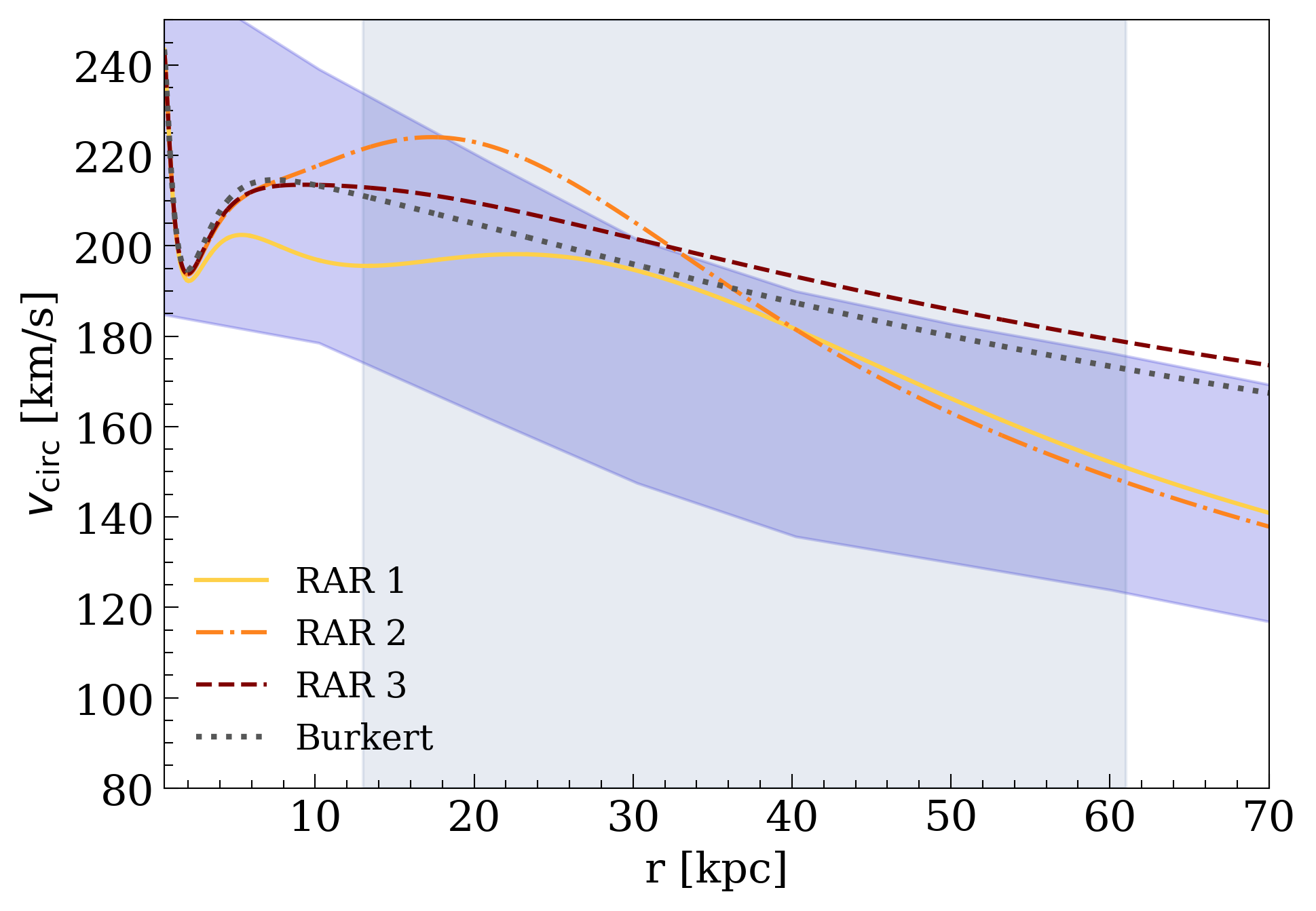}
    \caption{Circular velocity profiles for each DM model studied in this work. The shaded gray region represents the zone where the progenitor moves along its way around the Milky Way. The shaded blue region denotes the reproduced $95\%$ confidence region of the best-fit circular velocity curve of \cite{Gibbons2014}. Note that the four circular velocity curves agree relatively well with the best-fit prediction, despite slight differences at upper marginal values. At the same time, for $r \gtrsim 40$ kpc the polytropic profiles (RAR 1 and RAR 2) fall sharper than the power-law-like profiles (RAR 3 and Burkert), as expected based on Fig.~\ref{fig:mass_profiles_orbit_region}.}
    \label{fig:circular_velocities}
\end{figure}

As a complementary analysis, in Fig.~\ref{fig:circular_velocities} the four circular velocity profiles corresponding to each DM model are displayed. In spite of the fact that the four models behave differently from each other regarding the circular velocity (or similarly, the acceleration field at $z=0$), each one falls relatively well within the $95\%$ confidence region of the best-fit model of \cite{Gibbons2014}.

\begin{table}
    \caption{Stream morphological parameters.}
    \centering
    \begin{tabular}{cccc}
        \hline
        Model & $a_{L}$ [kpc] & $a_{T}$ [kpc] & $\psi$ [deg] \\
        \hline
        RAR 1 & $44.8$ & $66.9$ & $151.8$ \\
        RAR 2 & $43.9$ & $62.4$ & $136.4$ \\
        RAR 3 & $44.9$ & $76.0$ & $122.0$ \\
        Burkert & $44.4$ & $75.5$ & $120.3$ \\
        \hline
    \end{tabular}
    \tablefoot{The values are the Galactocentric apocentric distances of the leading and trailing tails, $a_{L}$ and $a_{T}$ respectively, in addition to the angular aperture between them ($\psi$) measured with respect to the heliocentric Sgr reference system. The apocentric distances correspond to the maximum value of the Gaussian function that delineates the shape of the stream tails.}
    \label{tab:apocenters}
\end{table}

\begin{figure*}
    \centering
    \includegraphics[width=0.495\textwidth]{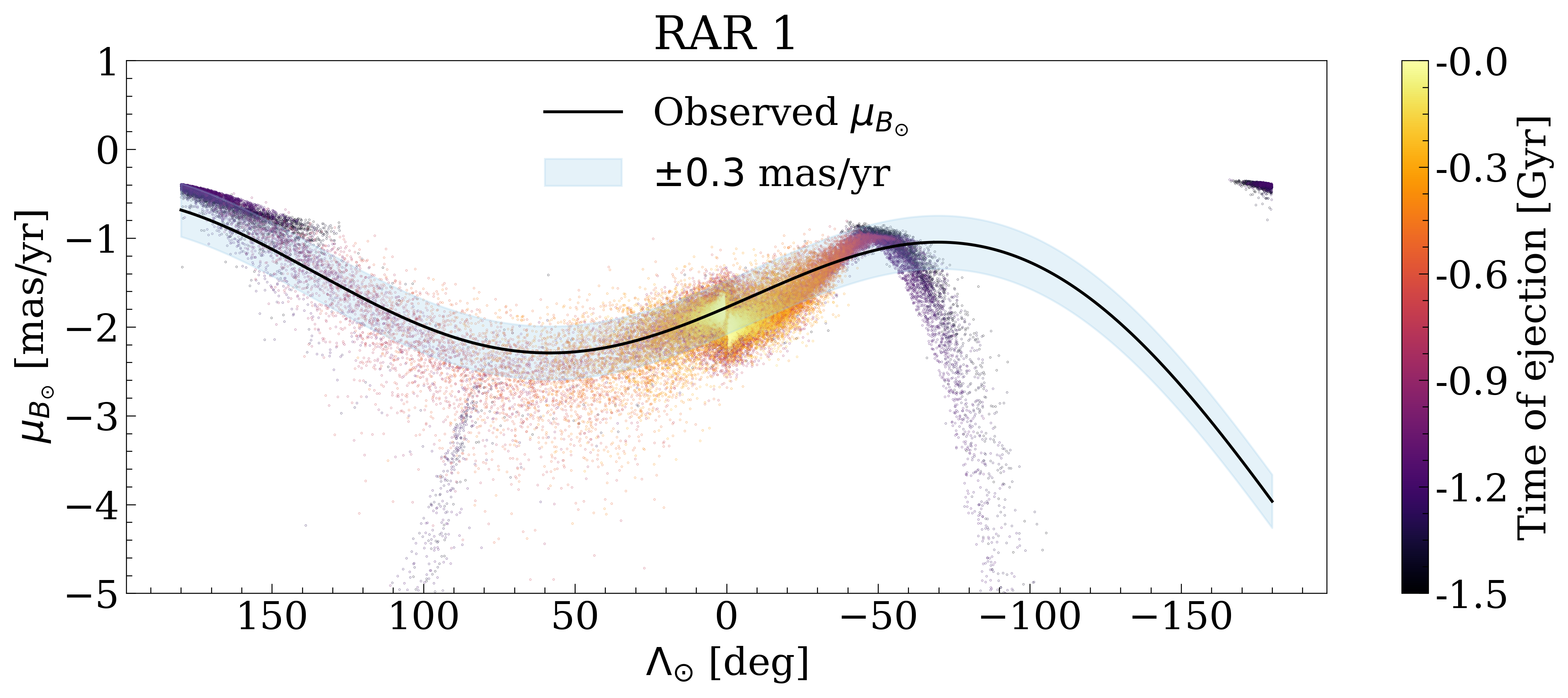}
    \hfill
    \includegraphics[width=0.495\textwidth]{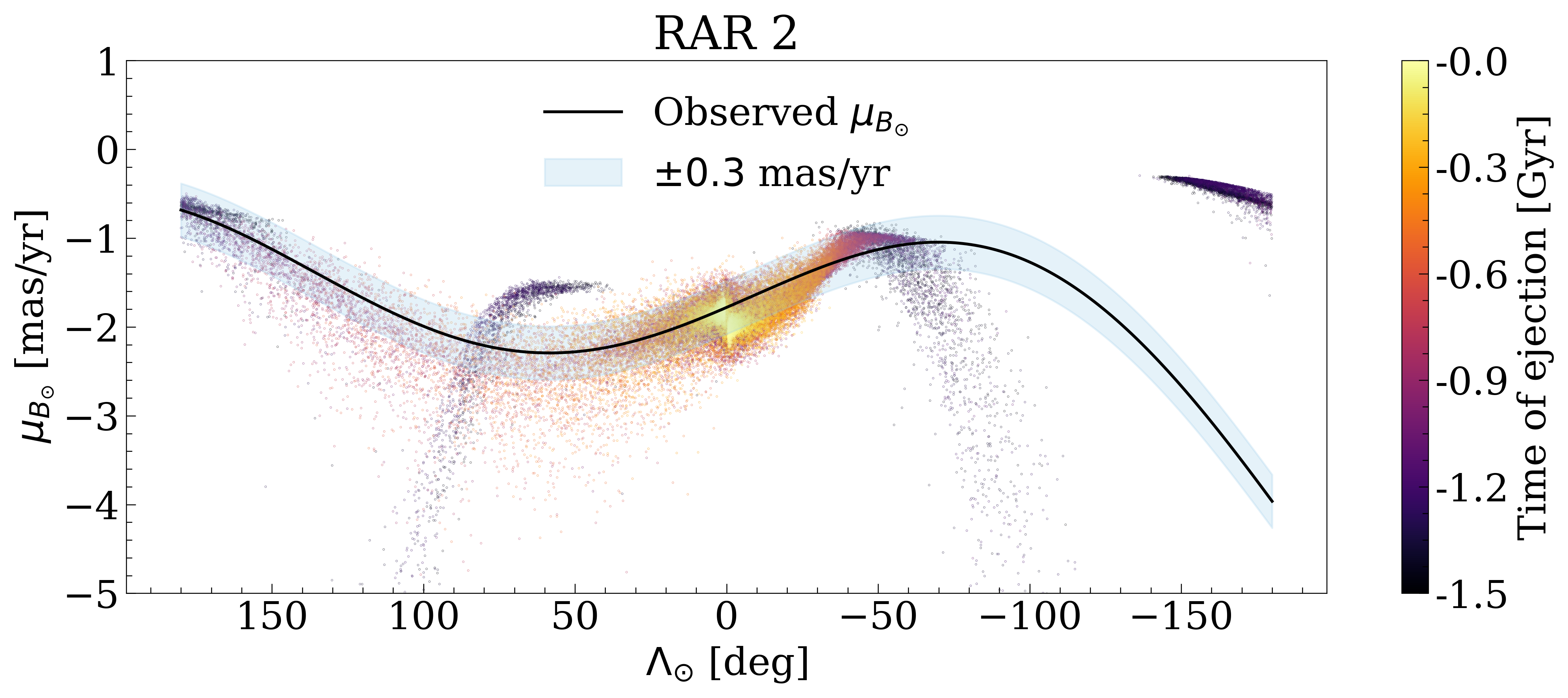}
    \vskip\baselineskip
    \includegraphics[width=0.495\textwidth]{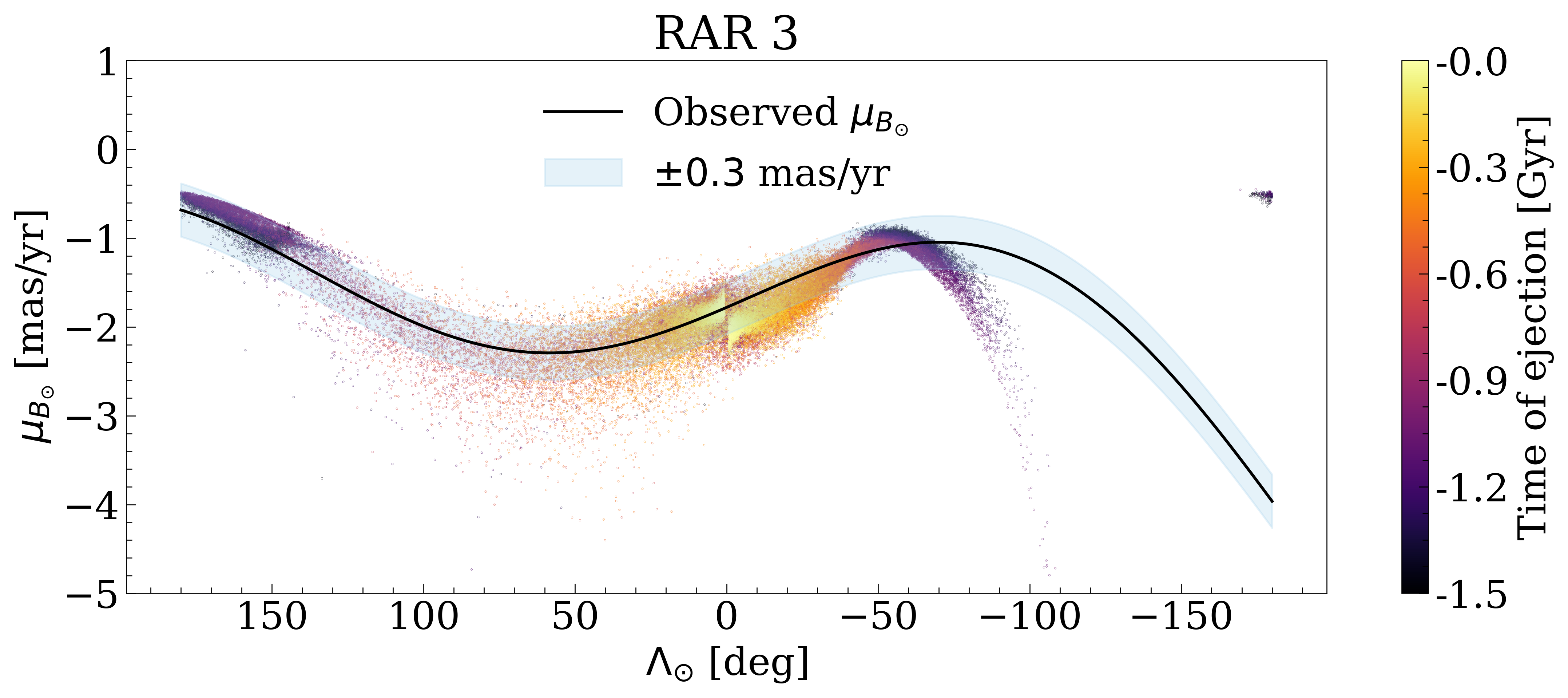}
    \hfill
    \includegraphics[width=0.495\textwidth]{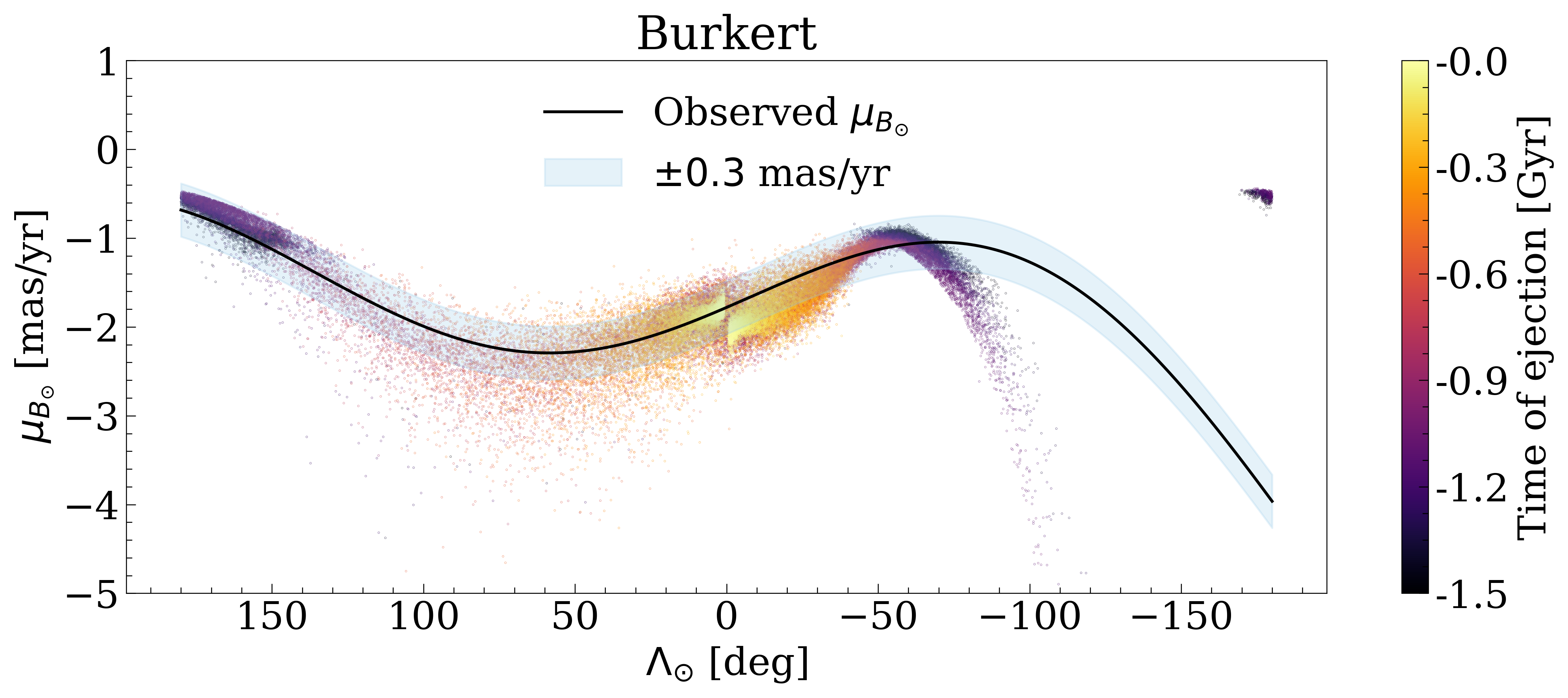}
    \caption{Mock stream star $\mu_{B_{\odot}}$ components at present, color coded by the ejection time. The solid black line represents the sigma-clipping fitted line computed by I$+$2020. The light-blue shaded region is the confidence interval shown in Fig.~\ref{fig:mu_vs_lambda}.}
    \label{fig:muB_vs_L}
\end{figure*}

\begin{figure*}
    \centering
    \includegraphics[width=0.495\textwidth]{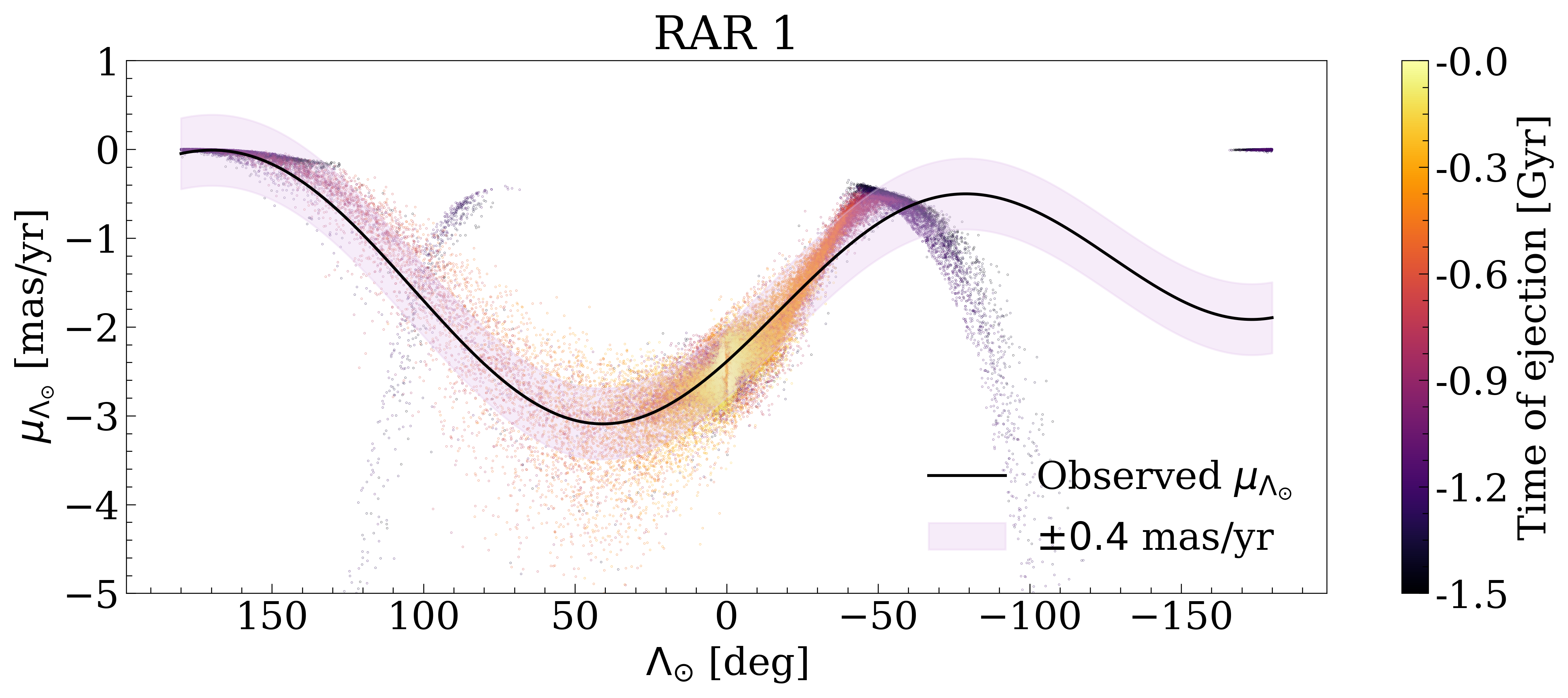}
    \hfill
    \includegraphics[width=0.495\textwidth]{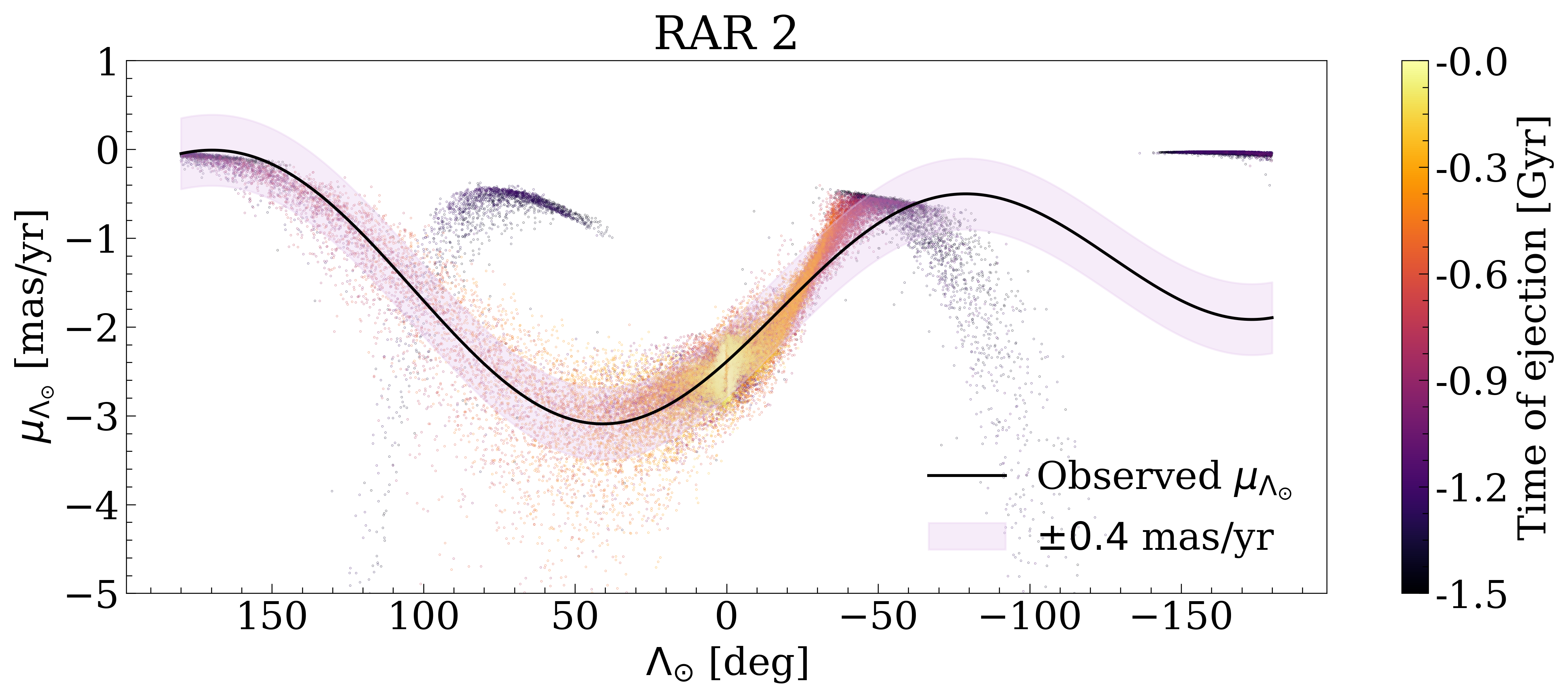}
    \vskip\baselineskip
    \includegraphics[width=0.495\textwidth]{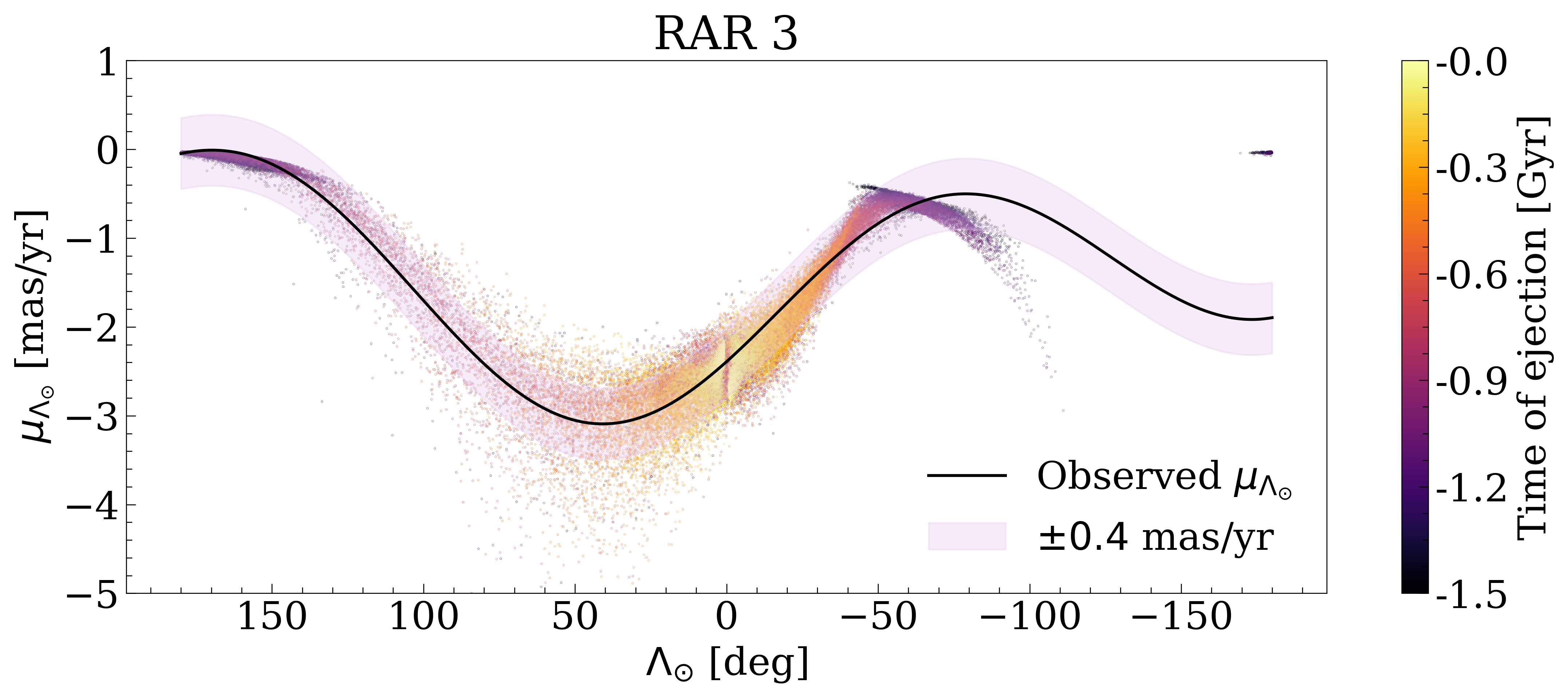}
    \hfill
    \includegraphics[width=0.495\textwidth]{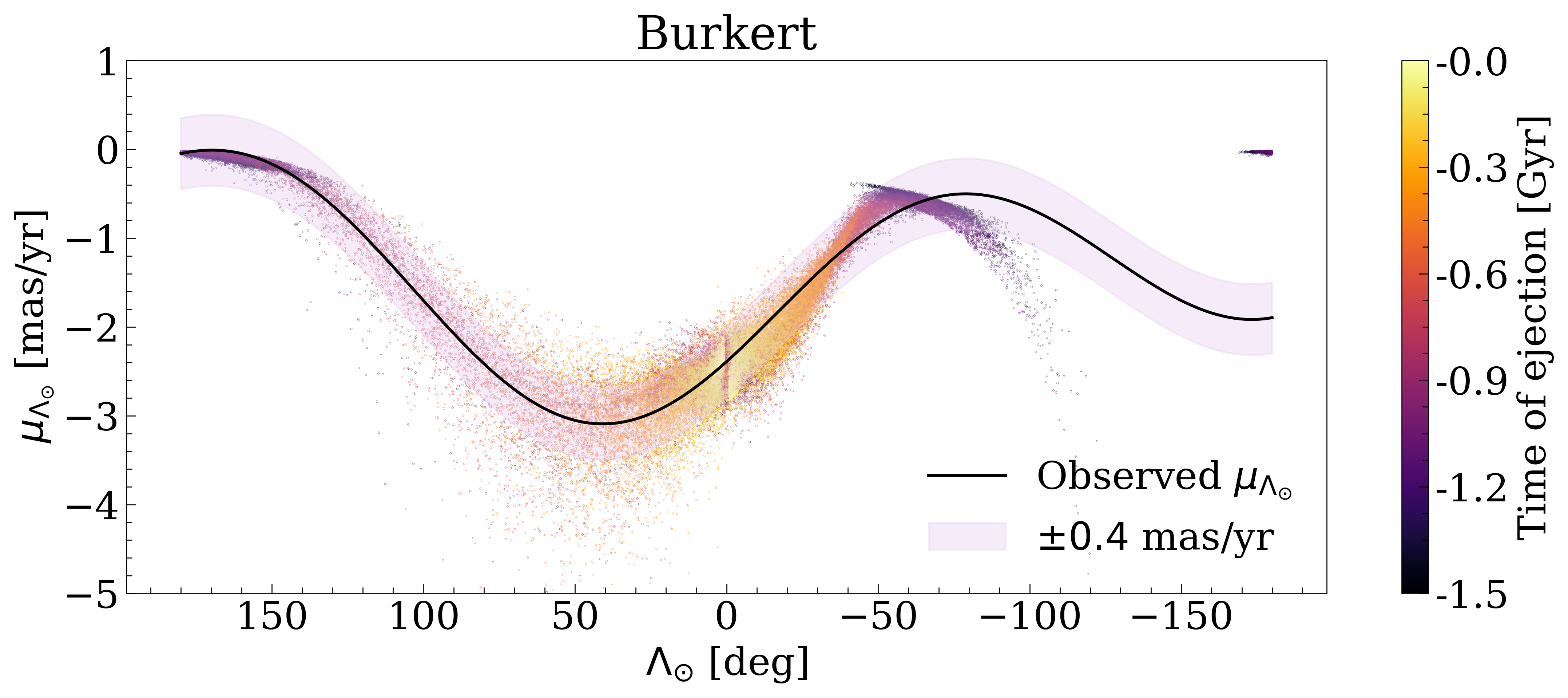}
    \caption{Same as in \cref{fig:muB_vs_L}, but for the $\mu_{\Lambda_{\odot}}$ component of the proper motion.}
    \label{fig:muL_vs_L}
\end{figure*}

\begin{figure*}
    \centering
    \includegraphics[width=0.495\textwidth]{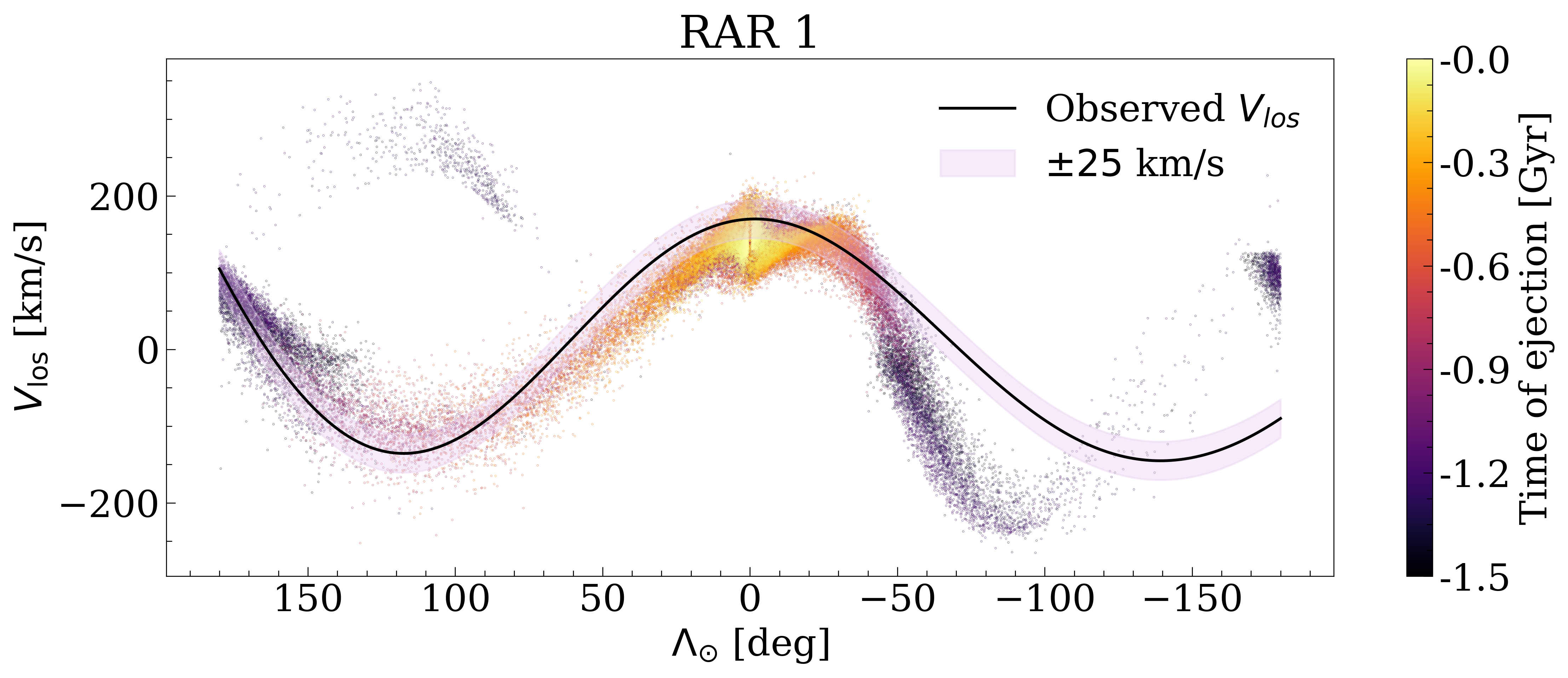}
    \hfill
    \includegraphics[width=0.495\textwidth]{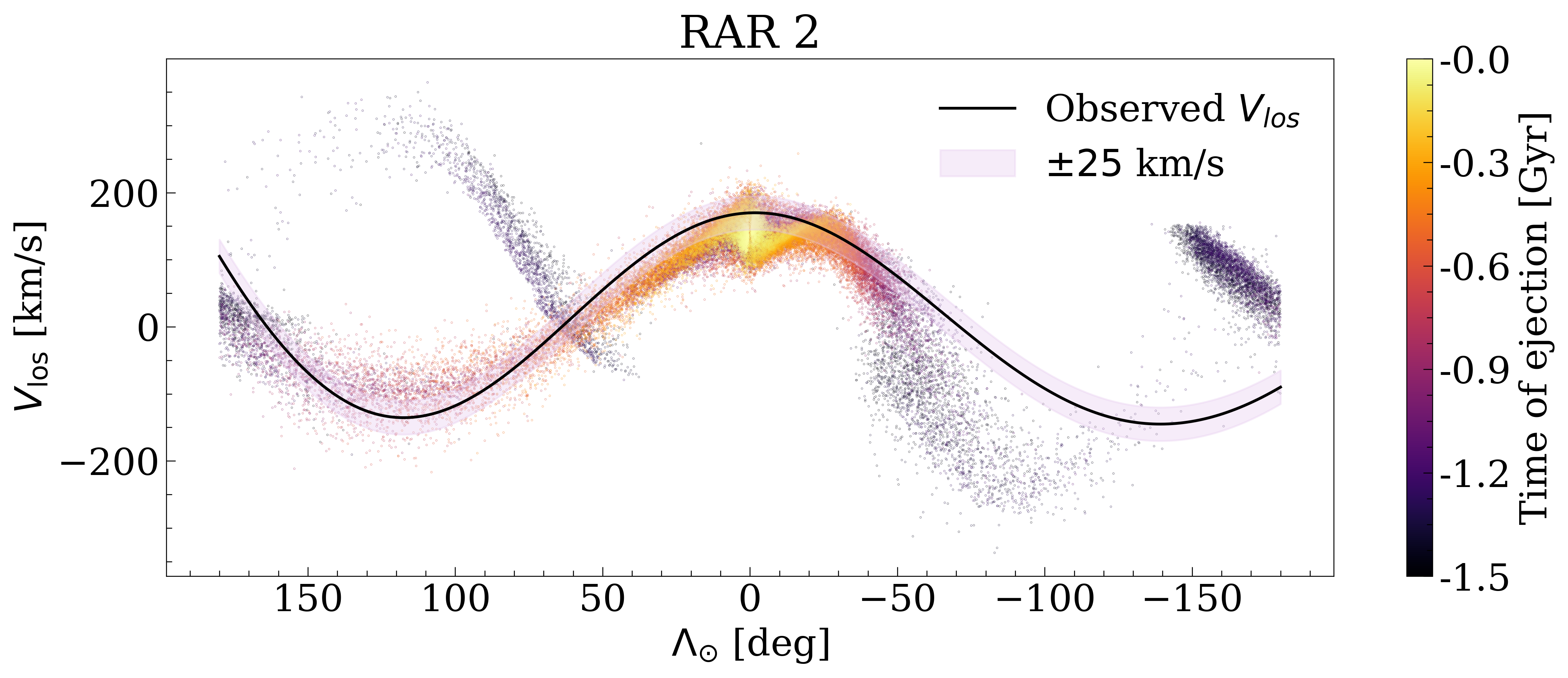}
    \vskip\baselineskip
    \includegraphics[width=0.495\textwidth]{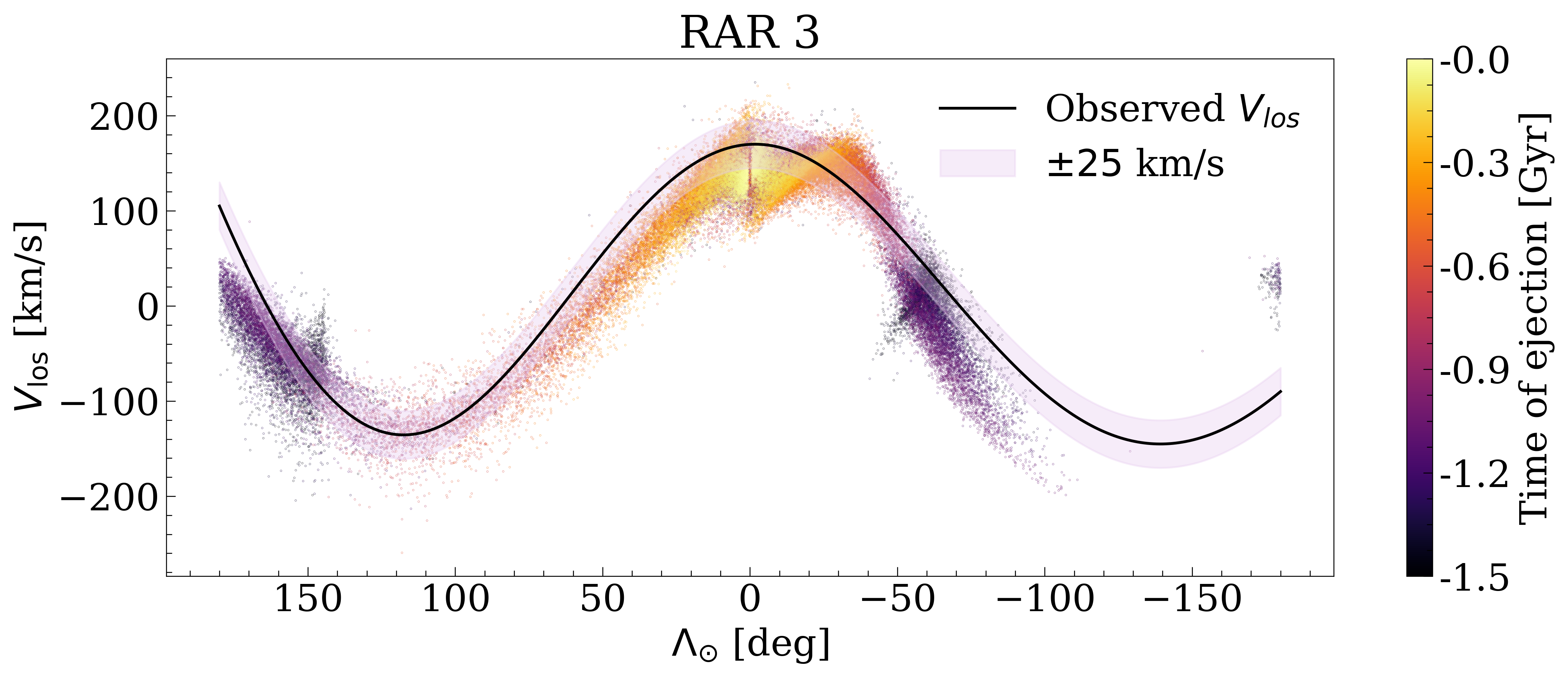}
    \hfill
    \includegraphics[width=0.495\textwidth]{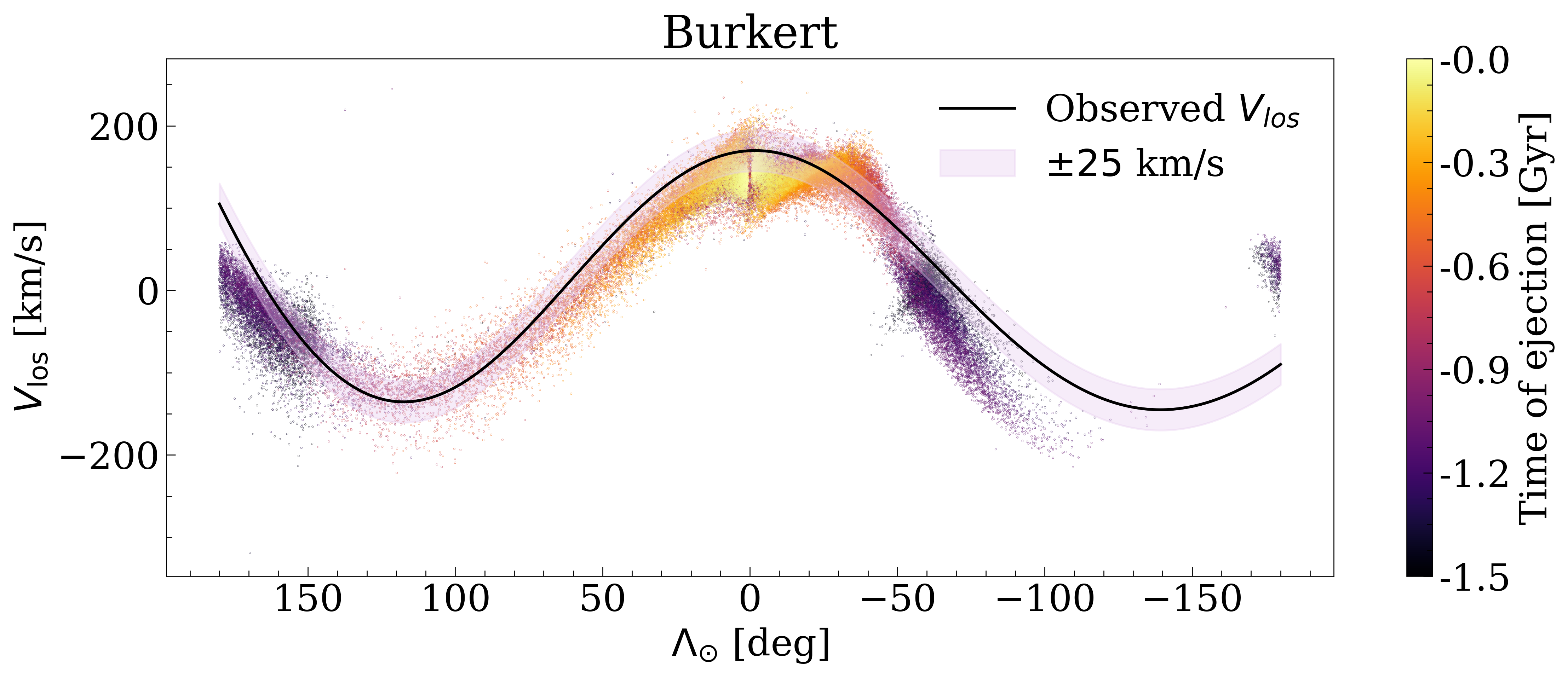}
    \caption{Line of sight velocity component of the stars of the four predicted streams at present, color coded with their ejection time. The solid black line represents the fitted line of sight velocity as a function of $\Lambda_{\odot}$ computed by I$+$2020, based on the \cite{Law2010} model. The grey shaded region represents the limits to account for Galactic field star contaminants, though here it was used a more stringent value than in I$+$2020 to reduce the contamination fraction.}
    \label{fig:losvel_vs_L}
\end{figure*}

We also compute the star distributions in the velocity space, showing the predicted stream in the $\Lambda_{\odot}-\mu_{B_{\odot}}$, $\Lambda_{\odot}-\mu_{\Lambda_{\odot}}$ and $\Lambda_{\odot}-V_{\mathrm{los}} $ planes, trying to reproduce the curves of Fig.~\ref{fig:mu_vs_lambda}. The result is shown respectively in Figs.~\ref{fig:muB_vs_L}, \ref{fig:muL_vs_L} and \ref{fig:losvel_vs_L}. In each one of the Figs.~\ref{fig:xz_time_of_ejection}, \ref{fig:dist_lambda}, \ref{fig:muB_vs_L}-\ref{fig:losvel_vs_L}, we display one plot for each of the four models studied. In addition, we computed the time of ejection of each of the stars, encoded in the color bar.

In Figs. \ref{fig:muB_vs_L} and \ref{fig:muL_vs_L}, the shaded region corresponds to a contamination of $11\%$ in the stream stars selected by I$+$2020. As can be seen, the trailing tail of the stellar stream resides in this region in each of the eight figures, but not the part of the stream with negative values of $\Lambda_{\odot}$. This section of the stream corresponds to the youngest piece of the leading tail. This suggests that the gravitational potentials tested here are unable to reproduce the entire observables independently of the type of halo tail considered (i.e. power-law-like or polytropic), or of the degree of DM mass concentration in the Galactocentric range of $10-60$ kpc. The same type of discrepant behavior is seen in Fig.~\ref{fig:losvel_vs_L}. This kind of disagreement is not new, and it is typical of DM models with spherical symmetry applied to model the Sgr stream \citep[see, e.g., discussions in][]{Helmi2004, Law2010}. However, the first-principle physics core--halo (RAR) profiles have never been tested before with the Sgr stream. Then, we have for the first time exploited the flexibility of such a fermionic model, showing the extent of success of the theory when confronted with full 6D phase-space data. It is only after this first step that one can start to properly explore further extensions allowed by the fermionic theory, such as the inclusion of the LMC or some degree of asphericity in the DM halo, left for future work. As an additional advantage, the dense fermion core developed in the cases with $m=56$ keV works as an alternative to the BH scenario by reproducing the orbits of the S-cluster stars~\citep{BecerraVergara2020,2021MNRAS.505L..64B}.

Comparing in more detail the degree of success between the models here studied, it is worth noting that the power law profiles (i.e. RAR $3$ and Burkert) achieve a slightly better agreement with the leading arm polynomial observables than the polytropic profiles (RAR $1$ and RAR $2$). Specifically, their distribution of stars in the leading tail for both proper motions and the line of sight velocity are less steep than in the polytropic cases. Although the extent of the leading arm is more sensitive to the accumulated mass in the first $10-60$ kpc, the above comparison between the RAR models suggests that higher values of halo mass within the $30-60$ kpc are preferable. On the other hand, the more extended power law density profiles with corresponding larger total halo masses with respect to the polytropic cases, seem not to improve in a considerable manner the trailing tail fitting.

Coming back to the discussion about spherically symmetric halos, evidence has suggested in the past that a spherical dark matter halo can not reproduce the entire set of observables. For example, in \cite{Helmi2004}, it is shown for the first time that modeling the dark halo of our Galaxy as a prolate one is the preferred option over other symmetries, for the region probed by the dynamically old debris (2-4 Gyr) from the leading arm. On the other hand, \cite{Johnston2005} have used M giant stars of tidal debris associated with Sgr to fit the simulated precession of the satellite's orbit. They have found that the best dark matter halo potential is one with slightly flattened potential contours. Thus, the difference with \cite{Helmi2004} is that they have determined that oblate dark halos are preferred over prolate ones. This is because the prolate halos can fit the Sgr leading arm radial velocities, but at the same time, they can not reproduce a stream fitting the precession of the progenitor, whereas oblate halos achieve the opposite.

This is why later on in \cite{Law2010}, an $N-$body model of the Sgr disruption was developed under the (ad-hoc) assumption of triaxiality for the dark halo, in order to better reproduce the stream. This non-axisymmetric potential fits the majority of angular position, distance, and radial velocity observables of the tidal debris of the progenitor. They determined that this triaxial component is a near oblate ellipsoid whose minor axis lies in the Galactic disk plane. This is an unlikely dynamical configuration, so they suggest that the orientation may have evolved with time or that the need for triaxiality may have been obviated by introducing another non-axisymmetric component. Such a component could be a time-dependent perturbation exerted by a massive satellite of the Milky Way, like the Magellanic Clouds.

Moreover, in a more sophisticated study carried out by \cite{Vasiliev2021}, they found a misalignment between the track of the stream and the direction of the reflex-corrected proper motions in the leading tail. They tested models without the presence of the LMC and found that these models can not solve the mentioned misalignment and, additionally, they overestimate the distance to the leading arm apocenter. In that same paper, they propose as a solution to this problem that there is a time-dependent perturbation acting on the whole stellar system. This perturbation may arise because of the presence of the LMC. They showed that the stream can be modeled up to a certain extent by taking into account the gravitational pull exerted by the LMC on our Galaxy (reflex motion) and on the Sgr system (dwarf and stream). They were able to provide a flexible mathematical function (depending on $11$ free parameters) for the DM halo, which is able to make predictions in agreement with all the available observations. The complexity in this kind of phenomenological modeling includes radially varying axis ratios and orientations. The inner part of it is oblate, with the minor axis lying perpendicular to the Galactic disk, while the outer one is another oblate ellipsoid whose minor axis lies almost in the Galactic plane. The degree of success of this model makes \cite{Vasiliev2021} to be a state-of-the-art description of the Sgr stellar stream.

\section{Conclusions}
\label{sec:final}

In this work, we have tested for the first time in the literature a fermionic dark matter halo model based on first-principle physics, such as (quantum) statistical mechanics and thermodynamics (i.e. the RAR distribution), against Sgr stellar stream observations. This is to try to reproduce the $6$D phase-space properties of the Sgr tidal stream with the aim of assessing the fermionic model. This self-gravitating system of neutral fermions has properties which make it a very good candidate to model galactic halos, as recently shown in \cite{2023ApJ...945....1K} for a sample of 120 different galaxies. Moreover, as demonstrated in \cite{2021MNRAS.502.4227A}, this fermionic core--halo profile with particle mass range in the sub-MeV range can arise at late stages of non-linear structure formation in cosmology, thus making it a promising case to be compared with other phenomenological profiles.

Using baryonic potentials in combination with RAR dark matter halos, we have modeled the gravitational potential for both the MW and the Sgr dwarf. After introducing a spray algorithm and a tidal radius model, we have applied them to create adequate initial conditions to integrate the orbit of individual stars in order to generate four stellar streams based on four different dark halo models. The data we used to compare with the theory is that of \cite{Ibata2020} and \cite{Vasiliev2021}: the cartesian Galactocentric coordinates of stars, where a distance trend was derived, and three polynomials functions taking into account the proper motions and line of sight velocity behavior of stars in the Sgr debris.

As the main conclusion, we find that across the different families of fermionic halo models (i.e. power-law-like and polytropic ones), they can only reproduce the trailing arm of the Sgr stream. Within the observationally allowed span of enclosed masses where the stream moves, none of the RAR halo profiles can answer for the observed trend of the leading tail. This is a conclusion which coincides with the results from other types of spherically symmetric haloes studied in the literature. Nevertheless, our spherically symmetric fermionic models do agree with the observed apocentric distance of the leading arm (see Fig.~\ref{fig:leading_and_trailing_apocenter_shapes}).

If one considers the previous success of this fermionic haloes when contrasted with different halo tracers of our Galaxy, such as the GD-1 cold stream~\citep{Mestre2024} or the ones associated with rotation curves analysis \citep{Arguelles2018,BecerraVergara2020,2023Univ....9..372A}, the more complex Sgr stream tracer here studied puts a firm limitation to the applicability of this first-principle physics theory on outer halo scales.

Indeed, the knowledge and tests done in this work can now be used to further extend the model by including the effects of the LMC and possibly add some degree of triaxiality, which may well be associated with an out-of-equilibrium DM component acquired within the merger history of the Galaxy. Further improvements to our theory may include more sophisticated models for stream generation, like the restricted N-body simulations developed in \cite{Vasiliev2021} or \cite{Ibata2024}. As already mentioned above for the Milky Way, but also extended now to other galaxy types, the motivation to maintain this kind of core--halo fermionic models in future more sophisticated analysis, is based on the success it has already achieved in explaining: (i) the rotation curve data of different galaxies \citep{Arguelles2019,2023ApJ...945....1K}; (ii) the universal galaxy scaling relations among different galaxy types \citep{Arguelles2019,2023ApJ...945....1K}; (iii) the formation and stability of fermionic profiles in a cosmological framework, as well as the existence of a critical point of DM core-collapse towards a supermassive BH \citep{2021MNRAS.502.4227A,2023MNRAS.523.2209A,2024ApJ...961L..10A}; (iv) the explanation of the Milky Way rotation curve and the phase-space track of the GD-1 stellar stream together with the astrometric data of the central S-stars (without assuming a BH), \citep{BecerraVergara2020,2021MNRAS.505L..64B,2022MNRAS.511L..35A,Mestre2024}; and (v) the prediction of the relativistic images and spectra around the central core when illuminated by an accretion disc \citep{2024MNRAS.534.1217P}.

In addition, we have consistently used the same fermion mass for both MW and Sgr dwarf halos. Moreover, the total enclosed mass of the dark halo of Sgr in the polytropic RAR profiles agrees with the estimates of the actual mass of the remnant discussed in \cite{Vasiliev2020a} and in \cite{Vasiliev2021}.

As a final remark, we stress that models depending on physically inspired free parameters are complementary to state-of-the-art MW halo models (which depend on several free parameters with consequent high degeneracy) and may help in the task of better reproducing the Sgr observations. This is why we believe that dark matter halo models like the one shown here are worth for further investigation in the field of stellar streams.

\begin{acknowledgements}
     The authors would like to thank Rodrigo Ibata for providing us with valuable observational data of the Sgr stream. We also thank Daniel Carpintero for fruitful discussions. SC acknowledges financial support from CONICET. MFM acknowledges support from CONICET (PIP2169) and from the Universidad Nacional de La Plata (PID G178). CRA acknowledges support from CONICET, the ANPCyT (grant PICT-2020-02990), and ICRANet. SC and CRA acknowledge support from Universidad Nacional de La Plata (PID G175).
\end{acknowledgements}

\bibliographystyle{aa}
\bibliography{refs}

\begin{thebibliography}{79}
\expandafter\ifx\csname natexlab\endcsname\relax\def\natexlab#1{#1}\fi

\bibitem[{{Antoja} {et~al.}(2020){Antoja}, {Ramos}, {Mateu}, {Helmi}, {Anders},
  {Jordi}, \& {Carballo-Bello}}]{Antoja2020}
{Antoja}, T., {Ramos}, P., {Mateu}, C., {et~al.} 2020, \aap, 635, L3

\bibitem[{{Arg{\"u}elles} {et~al.}(2023{\natexlab{a}}){Arg{\"u}elles},
  {Becerra-Vergara}, {Rueda}, \& {Ruffini}}]{2023Univ....9..197A}
{Arg{\"u}elles}, C.~R., {Becerra-Vergara}, E.~A., {Rueda}, J.~A., \& {Ruffini},
  R. 2023{\natexlab{a}}, Universe, 9, 197

\bibitem[{{Arg{\"u}elles} {et~al.}(2023{\natexlab{b}}){Arg{\"u}elles},
  {Boshkayev}, {Krut}, {Nurbakhyt}, {Rueda}, {Ruffini}, {Uribe-Su{\'a}rez}, \&
  {Yunis}}]{2023MNRAS.523.2209A}
{Arg{\"u}elles}, C.~R., {Boshkayev}, K., {Krut}, A., {et~al.}
  2023{\natexlab{b}}, \mnras, 523, 2209

\bibitem[{{Arg{\"u}elles} \& {Collazo}(2023)}]{2023Univ....9..372A}
{Arg{\"u}elles}, C.~R. \& {Collazo}, S. 2023, Universe, 9, 372

\bibitem[{{Arg{\"u}elles} {et~al.}(2021){Arg{\"u}elles}, {D{\'\i}az}, {Krut},
  \& {Yunis}}]{2021MNRAS.502.4227A}
{Arg{\"u}elles}, C.~R., {D{\'\i}az}, M.~I., {Krut}, A., \& {Yunis}, R. 2021,
  \mnras, 502, 4227

\bibitem[{{Arg{\"u}elles} {et~al.}(2022){Arg{\"u}elles}, {Mestre},
  {Becerra-Vergara}, {Crespi}, {Krut}, {Rueda}, \&
  {Ruffini}}]{2022MNRAS.511L..35A}
{Arg{\"u}elles}, C.~R., {Mestre}, M.~F., {Becerra-Vergara}, E.~A., {et~al.}
  2022, \mnras, 511, L35

\bibitem[{{Arg{\"u}elles} {et~al.}(2024){Arg{\"u}elles}, {Rueda}, \&
  {Ruffini}}]{2024ApJ...961L..10A}
{Arg{\"u}elles}, C.~R., {Rueda}, J.~A., \& {Ruffini}, R. 2024, \apjl, 961, L10

\bibitem[{Argüelles {et~al.}(2018)Argüelles, Krut, Rueda, \&
  Ruffini}]{Arguelles2018}
Argüelles, C., Krut, A., Rueda, J., \& Ruffini, R. 2018, Physics of the Dark
  Universe, 21, 82

\bibitem[{Argüelles {et~al.}(2019)Argüelles, Krut, Rueda, \&
  Ruffini}]{Arguelles2019}
Argüelles, C., Krut, A., Rueda, J., \& Ruffini, R. 2019, Physics of the Dark
  Universe, 24, 100278

\bibitem[{{Becerra-Vergara} {et~al.}(2020){Becerra-Vergara}, {Arg\"uelles},
  {Krut}, {Rueda}, \& {Ruffini}}]{BecerraVergara2020}
{Becerra-Vergara}, E.~A., {Arg\"uelles}, C.~R., {Krut}, A., {Rueda}, J.~A., \&
  {Ruffini}, R. 2020, A\&A, 641, A34

\bibitem[{{Becerra-Vergara} {et~al.}(2021){Becerra-Vergara}, {Arg{\"u}elles},
  {Krut}, {Rueda}, \& {Ruffini}}]{2021MNRAS.505L..64B}
{Becerra-Vergara}, E.~A., {Arg{\"u}elles}, C.~R., {Krut}, A., {Rueda}, J.~A.,
  \& {Ruffini}, R. 2021, \mnras, 505, L64

\bibitem[{Belokurov {et~al.}(2007)Belokurov, Evans, Irwin, Lynden-Bell, Yanny,
  Vidrih, Gilmore, Seabroke, Zucker, Wilkinson, Hewett, Bramich, Fellhauer,
  Newberg, Wyse, Beers, Bell, Barentine, Brinkmann, Cole, Pan, \&
  York}]{Belokurov2007}
Belokurov, V., Evans, N.~W., Irwin, M.~J., {et~al.} 2007, The Astrophysical
  Journal, 658, 337

\bibitem[{{Belokurov} {et~al.}(2014){Belokurov}, {Koposov}, {Evans},
  {Pe{\~n}arrubia}, {Irwin}, {Smith}, {Lewis}, {Gieles}, {Wilkinson},
  {Gilmore}, {Olszewski}, \& {Niederste-Ostholt}}]{Belokurov2014}
{Belokurov}, V., {Koposov}, S.~E., {Evans}, N.~W., {et~al.} 2014, \mnras, 437,
  116

\bibitem[{{Bonaca} \& {Hogg}(2018)}]{2018ApJ...867..101B}
{Bonaca}, A. \& {Hogg}, D.~W. 2018, \apj, 867, 101

\bibitem[{{Bonaca} \& {Price-Whelan}(2024)}]{Bonaca2024}
{Bonaca}, A. \& {Price-Whelan}, A.~M. 2024, arXiv e-prints, arXiv:2405.19410

\bibitem[{{Bowden} {et~al.}(2015){Bowden}, {Belokurov}, \&
  {Evans}}]{Bowden2015}
{Bowden}, A., {Belokurov}, V., \& {Evans}, N.~W. 2015, \mnras, 449, 1391

\bibitem[{{Burkert}(1995)}]{Burkert1995}
{Burkert}, A. 1995, \apjl, 447, L25

\bibitem[{{Chavanis}(2004)}]{Chavanis2}
{Chavanis}, P.-H. 2004, Physica A Statistical Mechanics and its Applications,
  332, 89

\bibitem[{{Debattista} {et~al.}(2013){Debattista}, {Ro{\v{s}}kar}, {Valluri},
  {Quinn}, {Moore}, \& {Wadsley}}]{Debattista2013}
{Debattista}, V.~P., {Ro{\v{s}}kar}, R., {Valluri}, M., {et~al.} 2013, \mnras,
  434, 2971

\bibitem[{{Deg} \& {Widrow}(2013)}]{Deg2013}
{Deg}, N. \& {Widrow}, L. 2013, \mnras, 428, 912

\bibitem[{{Dierickx} \& {Loeb}(2017)}]{Dierickx2017}
{Dierickx}, M. I.~P. \& {Loeb}, A. 2017, \apj, 836, 92

\bibitem[{Dinescu {et~al.}(2002)Dinescu, Majewski, Girard, Méndez, Sandage,
  Siegel, Kunkel, Subasavage, \& Ostheimer}]{Dinescu2002}
Dinescu, D.~I., Majewski, S.~R., Girard, T.~M., {et~al.} 2002, The
  Astrophysical Journal, 575, L67

\bibitem[{Fardal {et~al.}(2015)Fardal, Huang, \& Weinberg}]{Fardal2015}
Fardal, M.~A., Huang, S., \& Weinberg, M.~D. 2015, Monthly Notices of the Royal
  Astronomical Society, 452, 301

\bibitem[{{Fardal} {et~al.}(2019){Fardal}, {van der Marel}, {Law}, {Sohn},
  {Sesar}, {Hernitschek}, \& {Rix}}]{Fardal2019}
{Fardal}, M.~A., {van der Marel}, R.~P., {Law}, D.~R., {et~al.} 2019, \mnras,
  483, 4724

\bibitem[{{Gajda} \& {Lokas}(2016)}]{Gajda2016}
{Gajda}, G. \& {Lokas}, E.~L. 2016, \apj, 819, 20

\bibitem[{Gibbons {et~al.}(2014)Gibbons, Belokurov, \& Evans}]{Gibbons2014}
Gibbons, S. L.~J., Belokurov, V., \& Evans, N.~W. 2014, Monthly Notices of the
  Royal Astronomical Society, 445, 3788

\bibitem[{{GRAVITY Collaboration} {et~al.}(2018){GRAVITY Collaboration},
  {Abuter, R.}, {Amorim, A.}, {Anugu, N.}, {Baub\"ock, M.}, {Benisty, M.},
  {Berger, J. P.}, {Blind, N.}, {Bonnet, H.}, {Brandner, W.}, {Buron, A.},
  {Collin, C.}, {Chapron, F.}, {Cl\'enet, Y.}, {dCoud\'e u Foresto, V.}, {de
  Zeeuw, P. T.}, {Deen, C.}, {Delplancke-Str\"obele, F.}, {Dembet, R.},
  {Dexter, J.}, {Duvert, G.}, {Eckart, A.}, {Eisenhauer, F.}, {Finger, G.},
  {F\"orster Schreiber, N. M.}, {F\'edou, P.}, {Garcia, P.}, {Garcia Lopez,
  R.}, {Gao, F.}, {Gendron, E.}, {Genzel, R.}, {Gillessen, S.}, {Gordo, P.},
  {Habibi, M.}, {Haubois, X.}, {Haug, M.}, {Hau\ss{}mann, F.}, {Henning, Th.},
  {Hippler, S.}, {Horrobin, M.}, {Hubert, Z.}, {Hubin, N.}, {Jimenez Rosales,
  A.}, {Jochum, L.}, {Jocou, L.}, {Kaufer, A.}, {Kellner, S.}, {Kendrew, S.},
  {Kervella, P.}, {Kok, Y.}, {Kulas, M.}, {Lacour, S.}, {Lapeyr\`ere, V.},
  {Lazareff, B.}, {Le Bouquin, J.-B.}, {L\'ena, P.}, {Lippa, M.}, {Lenzen, R.},
  {M\'erand, A.}, {M\"uler, E.}, {Neumann, U.}, {Ott, T.}, {Palanca, L.},
  {Paumard, T.}, {Pasquini, L.}, {Perraut, K.}, {Perrin, G.}, {Pfuhl, O.},
  {Plewa, P. M.}, {Rabien, S.}, {Ram\'{\i}rez, A.}, {Ramos, J.}, {Rau, C.},
  {Rodr\'{\i}guez-Coira, G.}, {Rohloff, R.-R.}, {Rousset, G.},
  {Sanchez-Bermudez, J.}, {Scheithauer, S.}, {Sch\"oller, M.}, {Schuler, N.},
  {Spyromilio, J.}, {Straub, O.}, {Straubmeier, C.}, {Sturm, E.}, {Tacconi, L.
  J.}, {Tristram, K. R. W.}, {Vincent, F.}, {von Fellenberg, S.}, {Wank, I.},
  {Waisberg, I.}, {Widmann, F.}, {Wieprecht, E.}, {Wiest, M.}, {Wiezorrek, E.},
  {Woillez, J.}, {Yazici, S.}, {Ziegler, D.}, \& {Zins,
  G.}}]{GravityCollaboration2018}
{GRAVITY Collaboration}, {Abuter, R.}, {Amorim, A.}, {et~al.} 2018, A\&A, 615,
  L15

\bibitem[{Grillmair \& Dionatos(2006)}]{Grillmair_2006}
Grillmair, C.~J. \& Dionatos, O. 2006, The Astrophysical Journal, 643, L17

\bibitem[{{Hawkins} {et~al.}(2023){Hawkins}, {Price-Whelan}, {Sheffield},
  {Subrahimovic}, {Beaton}, {Belokurov}, {Erkal}, {Koposov}, {Lane}, {Laporte},
  \& {Nitschelm}}]{Hawkins2023}
{Hawkins}, K., {Price-Whelan}, A.~M., {Sheffield}, A.~A., {et~al.} 2023, \apj,
  948, 123

\bibitem[{{Helmi}(2004)}]{Helmi2004}
{Helmi}, A. 2004, \apjl, 610, L97

\bibitem[{{Hernitschek} {et~al.}(2017){Hernitschek}, {Sesar}, {Rix},
  {Belokurov}, {Martinez-Delgado}, {Martin}, {Kaiser}, {Hodapp}, {Chambers},
  {Wainscoat}, {Magnier}, {Kudritzki}, {Metcalfe}, \&
  {Draper}}]{Hernitschek2017}
{Hernitschek}, N., {Sesar}, B., {Rix}, H.-W., {et~al.} 2017, \apj, 850, 96

\bibitem[{{Huxor} \& {Grebel}(2015)}]{Huxor2015}
{Huxor}, A.~P. \& {Grebel}, E.~K. 2015, \mnras, 453, 2653

\bibitem[{Ibata {et~al.}(2020)Ibata, Bellazzini, Thomas, Malhan, Martin,
  Famaey, \& Siebert}]{Ibata2020}
Ibata, R., Bellazzini, M., Thomas, G., {et~al.} 2020, The Astrophysical
  Journal, 891, L19

\bibitem[{{Ibata} {et~al.}(2001){Ibata}, {Lewis}, {Irwin}, {Totten}, \&
  {Quinn}}]{Ibata2001}
{Ibata}, R., {Lewis}, G.~F., {Irwin}, M., {Totten}, E., \& {Quinn}, T. 2001,
  \apj, 551, 294

\bibitem[{{Ibata} {et~al.}(2024){Ibata}, {Malhan}, {Tenachi},
  {Ardern-Arentsen}, {Bellazzini}, {Bianchini}, {Bonifacio}, {Caffau},
  {Diakogiannis}, {Errani}, {Famaey}, {Ferrone}, {Martin}, {di Matteo},
  {Monari}, {Renaud}, {Starkenburg}, {Thomas}, {Viswanathan}, \&
  {Yuan}}]{Ibata2024}
{Ibata}, R., {Malhan}, K., {Tenachi}, W., {et~al.} 2024, The Astrophysical
  Journal, 967, 89

\bibitem[{{Ibata} {et~al.}(1994){Ibata}, {Gilmore}, \& {Irwin}}]{Ibata1994}
{Ibata}, R.~A., {Gilmore}, G., \& {Irwin}, M.~J. 1994, \nat, 370, 194

\bibitem[{{Ibata} {et~al.}(1997){Ibata}, {Wyse}, {Gilmore}, {Irwin}, \&
  {Suntzeff}}]{Ibata1997a}
{Ibata}, R.~A., {Wyse}, R. F.~G., {Gilmore}, G., {Irwin}, M.~J., \& {Suntzeff},
  N.~B. 1997, \aj, 113, 634

\bibitem[{{Johnston} {et~al.}(2005){Johnston}, {Law}, \&
  {Majewski}}]{Johnston2005}
{Johnston}, K.~V., {Law}, D.~R., \& {Majewski}, S.~R. 2005, \apj, 619, 800

\bibitem[{{Koposov} {et~al.}(2013){Koposov}, {Belokurov}, \&
  {Evans}}]{Koposov2013}
{Koposov}, S.~E., {Belokurov}, V., \& {Evans}, N.~W. 2013, \apj, 766, 79

\bibitem[{Koposov {et~al.}(2012)Koposov, Belokurov, Evans, Gilmore, Gieles,
  Irwin, Lewis, Niederste-Ostholt, Pe{\~{n}}arrubia, Smith, Bizyaev,
  Malanushenko, Malanushenko, Schneider, \& Wyse}]{Koposov2012}
Koposov, S.~E., Belokurov, V., Evans, N.~W., {et~al.} 2012, The Astrophysical
  Journal, 750, 80

\bibitem[{{Koposov} {et~al.}(2019){Koposov}, {Belokurov}, {Li}, {Mateu},
  {Erkal}, {Grillmair}, {Hendel}, {Price-Whelan}, {Laporte}, {Hawkins}, {Sohn},
  {del Pino}, {Evans}, {Slater}, {Kallivayalil}, {Navarro}, \& {Orphan Aspen
  Treasury Collaboration}}]{Koposov2019}
{Koposov}, S.~E., {Belokurov}, V., {Li}, T.~S., {et~al.} 2019, \mnras, 485,
  4726

\bibitem[{{Krut} {et~al.}(2023){Krut}, {Arg{\"u}elles}, {Chavanis}, {Rueda}, \&
  {Ruffini}}]{2023ApJ...945....1K}
{Krut}, A., {Arg{\"u}elles}, C.~R., {Chavanis}, P.~H., {Rueda}, J.~A., \&
  {Ruffini}, R. 2023, \apj, 945, 1

\bibitem[{{K{\"u}pper} {et~al.}(2012){K{\"u}pper}, {Lane}, \&
  {Heggie}}]{Kupper2012}
{K{\"u}pper}, A. H.~W., {Lane}, R.~R., \& {Heggie}, D.~C. 2012, \mnras, 420,
  2700

\bibitem[{{Law} {et~al.}(2005){Law}, {Johnston}, \& {Majewski}}]{Law2005}
{Law}, D.~R., {Johnston}, K.~V., \& {Majewski}, S.~R. 2005, The Astrophysical
  Journal, 619, 807

\bibitem[{Law \& Majewski(2010)}]{Law2010}
Law, D.~R. \& Majewski, S.~R. 2010, The Astrophysical Journal, 714, 229

\bibitem[{{Law} {et~al.}(2009){Law}, {Majewski}, \& {Johnston}}]{Law2009}
{Law}, D.~R., {Majewski}, S.~R., \& {Johnston}, K.~V. 2009, \apjl, 703, L67

\bibitem[{{Li} {et~al.}(2019){Li}, {Koposov}, {Zucker}, {Lewis}, {Kuehn},
  {Simpson}, {Ji}, {Shipp}, {Mao}, {Geha}, {Pace}, {Mackey}, {Allam}, {Tucker},
  {Da Costa}, {Erkal}, {Simon}, {Mould}, {Martell}, {Wan}, {De Silva},
  {Bechtol}, {Balbinot}, {Belokurov}, {Bland-Hawthorn}, {Casey}, {Cullinane},
  {Drlica-Wagner}, {Sharma}, {Vivas}, {Wechsler}, {Yanny}, \& {S5
  Collaboration}}]{Li2019}
{Li}, T.~S., {Koposov}, S.~E., {Zucker}, D.~B., {et~al.} 2019, \mnras, 490,
  3508

\bibitem[{{Luque} {et~al.}(2017){Luque}, {Pieres}, {Santiago}, {Yanny},
  {Vivas}, {Queiroz}, {Drlica-Wagner}, {Morganson}, {Balbinot}, {Marshall},
  {Li}, {Neto}, {da Costa}, {Maia}, {Bechtol}, {Kim}, {Bernstein}, {Dodelson},
  {Whiteway}, {Diehl}, {Finley}, {Abbott}, {Abdalla}, {Allam}, {Annis},
  {Benoit-L{\'e}vy}, {Bertin}, {Brooks}, {Burke}, {Rosell}, {Kind},
  {Carretero}, {Cunha}, {D'Andrea}, {Desai}, {Doel}, {Evrard}, {Flaugher},
  {Fosalba}, {Gerdes}, {Goldstein}, {Gruen}, {Gruendl}, {Gutierrez}, {James},
  {Kuehn}, {Kuropatkin}, {Lahav}, {Martini}, {Miquel}, {Nord}, {Ogando},
  {Plazas}, {Romer}, {Sanchez}, {Scarpine}, {Schubnell}, {Sevilla-Noarbe},
  {Smith}, {Soares-Santos}, {Sobreira}, {Suchyta}, {Swanson}, {Tarle},
  {Thomas}, \& {Walker}}]{Luque2017}
{Luque}, E., {Pieres}, A., {Santiago}, B., {et~al.} 2017, \mnras, 468, 97

\bibitem[{{Majewski} {et~al.}(2004){Majewski}, {Kunkel}, {Law}, {Patterson},
  {Polak}, {Rocha-Pinto}, {Crane}, {Frinchaboy}, {Hummels}, {Johnston}, {Rhee},
  {Skrutskie}, \& {Weinberg}}]{Majewski_2004AJ}
{Majewski}, S.~R., {Kunkel}, W.~E., {Law}, D.~R., {et~al.} 2004, \aj, 128, 245

\bibitem[{{Majewski} {et~al.}(1999){Majewski}, {Siegel}, {Kunkel}, {Reid},
  {Johnston}, {Thompson}, {Landolt}, \& {Palma}}]{Majewski1999}
{Majewski}, S.~R., {Siegel}, M.~H., {Kunkel}, W.~E., {et~al.} 1999, \aj, 118,
  1709

\bibitem[{Majewski {et~al.}(2003)Majewski, Skrutskie, Weinberg, \&
  Ostheimer}]{Majewski2003}
Majewski, S.~R., Skrutskie, M.~F., Weinberg, M.~D., \& Ostheimer, J.~C. 2003,
  The Astrophysical Journal, 599, 1082

\bibitem[{Malhan \& Ibata(2018)}]{STREAMFINDER2018}
Malhan, K. \& Ibata, R.~A. 2018, Monthly Notices of the Royal Astronomical
  Society, 477, 4063

\bibitem[{Mateo {et~al.}(1998)Mateo, Olszewski, \& Morrison}]{Mateo1998}
Mateo, M., Olszewski, E.~W., \& Morrison, H.~L. 1998, The Astrophysical
  Journal, 508, L55

\bibitem[{{Mateu}(2023)}]{Mateu2023}
{Mateu}, C. 2023, Monthly Notices of the Royal Astronomical Society, 520, 5225

\bibitem[{{Mestre} {et~al.}(2024){Mestre}, {Arg{\"u}elles}, {Carpintero},
  {Crespi}, \& {Krut}}]{Mestre2024}
{Mestre}, M.~F., {Arg{\"u}elles}, C.~R., {Carpintero}, D.~D., {Crespi}, V., \&
  {Krut}, A. 2024, \aap, 689, A194

\bibitem[{{Minniti} {et~al.}(2011){Minniti}, {Saito}, {Alonso-Garc{\'\i}a},
  {Lucas}, \& {Hempel}}]{Minniti2011}
{Minniti}, D., {Saito}, R.~K., {Alonso-Garc{\'\i}a}, J., {Lucas}, P.~W., \&
  {Hempel}, M. 2011, \apjl, 733, L43

\bibitem[{{Miyamoto} \& {Nagai}(1975)}]{Miyamoto1975}
{Miyamoto}, M. \& {Nagai}, R. 1975, \pasj, 27, 533

\bibitem[{{Newberg} {et~al.}(2007){Newberg}, {Yanny}, {Cole}, {Beers}, {Re
  Fiorentin}, {Schneider}, \& {Wilhelm}}]{Newberg2007}
{Newberg}, H.~J., {Yanny}, B., {Cole}, N., {et~al.} 2007, \apj, 668, 221

\bibitem[{{Newberg} {et~al.}(2003){Newberg}, {Yanny}, {Grebel}, {Hennessy},
  {Ivezi{\'c}}, {Martinez-Delgado}, {Odenkirchen}, {Rix}, {Brinkmann}, {Lamb},
  {Schneider}, \& {York}}]{Newberg2003}
{Newberg}, H.~J., {Yanny}, B., {Grebel}, E.~K., {et~al.} 2003, \apj, 596, L191

\bibitem[{{Newberg} {et~al.}(2002){Newberg}, {Yanny}, {Rockosi}, {Grebel},
  {Rix}, {Brinkmann}, {Csabai}, {Hennessy}, {Hindsley}, {Ibata}, {Ivezi{\'c}},
  {Lamb}, {Nash}, {Odenkirchen}, {Rave}, {Schneider}, {Smith}, {Stolte}, \&
  {York}}]{Newberg2002}
{Newberg}, H.~J., {Yanny}, B., {Rockosi}, C., {et~al.} 2002, \apj, 569, 245

\bibitem[{{Pelle} {et~al.}(2024){Pelle}, {Arg{\"u}elles}, {Vieyro}, {Crespi},
  {Millauro}, {Mestre}, {Reula}, \& {Carrasco}}]{2024MNRAS.534.1217P}
{Pelle}, J., {Arg{\"u}elles}, C.~R., {Vieyro}, F.~L., {et~al.} 2024, \mnras,
  534, 1217

\bibitem[{{Pouliasis} {et~al.}(2017){Pouliasis}, {Di Matteo}, \&
  {Haywood}}]{Pouliasis2017}
{Pouliasis}, E., {Di Matteo}, P., \& {Haywood}, M. 2017, A\&A, 598, A66

\bibitem[{{Ramos} {et~al.}(2020){Ramos}, {Mateu}, {Antoja}, {Helmi},
  {Castro-Ginard}, {Balbinot}, \& {Carrasco}}]{Ramos2020}
{Ramos}, P., {Mateu}, C., {Antoja}, T., {et~al.} 2020, \aap, 638, A104

\bibitem[{{Reid} {et~al.}(2014){Reid}, {Menten}, {Brunthaler}, {Zheng}, {Dame},
  {Xu}, {Wu}, {Zhang}, {Sanna}, {Sato}, {Hachisuka}, {Choi}, {Immer},
  {Moscadelli}, {Rygl}, \& {Bartkiewicz}}]{Reid2014}
{Reid}, M.~J., {Menten}, K.~M., {Brunthaler}, A., {et~al.} 2014, \apj, 783, 130

\bibitem[{Robles {et~al.}(2015)Robles, Lora, Matos, \&
  Sánchez-Salcedo}]{Robles2015}
Robles, V.~H., Lora, V., Matos, T., \& Sánchez-Salcedo, F.~J. 2015, The
  Astrophysical Journal, 810, 99

\bibitem[{{Ruffini} {et~al.}(2015){Ruffini}, {Arg{\"u}elles}, \&
  {Rueda}}]{Ruffini2015}
{Ruffini}, R., {Arg{\"u}elles}, C.~R., \& {Rueda}, J.~A. 2015, \mnras, 451, 622

\bibitem[{{Sale} {et~al.}(2010){Sale}, {Drew}, {Knigge}, {Zijlstra}, {Irwin},
  {Morris}, {Phillipps}, {Drake}, {Greimel}, {Unruh}, {Groot}, {Mampaso}, \&
  {Walton}}]{Sale2010}
{Sale}, S.~E., {Drew}, J.~E., {Knigge}, C., {et~al.} 2010, \mnras, 402, 713

\bibitem[{{Sch{\"o}nrich} {et~al.}(2010){Sch{\"o}nrich}, {Binney}, \&
  {Dehnen}}]{Schonrich2010}
{Sch{\"o}nrich}, R., {Binney}, J., \& {Dehnen}, W. 2010, \mnras, 403, 1829

\bibitem[{{Slater} {et~al.}(2013){Slater}, {Bell}, {Schlafly}, {Juri{\'c}},
  {Martin}, {Rix}, {Bernard}, {Burgett}, {Chambers}, {Finkbeiner}, {Goldman},
  {Kaiser}, {Magnier}, {Morganson}, {Price}, \& {Tonry}}]{Slater2013}
{Slater}, C.~T., {Bell}, E.~F., {Schlafly}, E.~F., {et~al.} 2013, \apj, 762, 6

\bibitem[{Sofue(2013)}]{Sofue2013}
Sofue, Y. 2013, Publications of the Astronomical Society of Japan, 65, 118

\bibitem[{{Springel}(2005)}]{Springel2005}
{Springel}, V. 2005, \mnras, 364, 1105

\bibitem[{Totten \& Irwin(1998)}]{Totten1998}
Totten, E.~J. \& Irwin, M.~J. 1998, Monthly Notices of the Royal Astronomical
  Society, 294, 1

\bibitem[{Varghese {et~al.}(2011)Varghese, Ibata, \& Lewis}]{Varghese2011}
Varghese, A., Ibata, R., \& Lewis, G.~F. 2011, Monthly Notices of the Royal
  Astronomical Society, 417, 198

\bibitem[{Vasiliev \& Belokurov(2020)}]{Vasiliev2020a}
Vasiliev, E. \& Belokurov, V. 2020, Monthly Notices of the Royal Astronomical
  Society, 497, 4162

\bibitem[{{Vasiliev} {et~al.}(2021){Vasiliev}, {Belokurov}, \&
  {Erkal}}]{Vasiliev2021}
{Vasiliev}, E., {Belokurov}, V., \& {Erkal}, D. 2021, \mnras, 501, 2279

\bibitem[{Virtanen {et~al.}(2020)Virtanen, Gommers, Oliphant, Haberland, Reddy,
  Cournapeau, Burovski, Peterson, Weckesser, Bright, {van der Walt}, Brett,
  Wilson, Millman, Mayorov, Nelson, Jones, Kern, Larson, Carey, Polat, Feng,
  Moore, {VanderPlas}, Laxalde, Perktold, Cimrman, Henriksen, Quintero, Harris,
  Archibald, Ribeiro, Pedregosa, {van Mulbregt}, \& {SciPy 1.0
  Contributors}}]{2020SciPy-NMeth}
Virtanen, P., Gommers, R., Oliphant, T.~E., {et~al.} 2020, Nature Methods, 17,
  261

\bibitem[{Walker {et~al.}(2009)Walker, Mateo, Olszewski, Pe{\~{n}}arrubia,
  Evans, \& Gilmore}]{Walker2009}
Walker, M.~G., Mateo, M., Olszewski, E.~W., {et~al.} 2009, The Astrophysical
  Journal, 704, 1274

\bibitem[{{Wang} {et~al.}(2022){Wang}, {Hammer}, {Yang}, \&
  {Wang}}]{2022ApJ...940L...3W}
{Wang}, H.-F., {Hammer}, F., {Yang}, Y.-B., \& {Wang}, J.-L. 2022, \apjl, 940,
  L3

\bibitem[{{Yanny} {et~al.}(2009){Yanny}, {Newberg}, {Johnson}, {Lee}, {Beers},
  {Bizyaev}, {Brewington}, {Fiorentin}, {Harding}, {Malanushenko},
  {Malanushenko}, {Oravetz}, {Pan}, {Simmons}, \& {Snedden}}]{Yanny2009}
{Yanny}, B., {Newberg}, H.~J., {Johnson}, J.~A., {et~al.} 2009, \apj, 700, 1282

\end{thebibliography}

\end{document}